\documentclass[11pt]{article}

\usepackage[textwidth=15.5cm,textheight=22.5cm]{geometry}
\usepackage{amsmath,amssymb}
\usepackage{latexsym}
\usepackage{graphicx}
\usepackage{cite}
\usepackage{subfig}
\usepackage{bbm}
\usepackage{bm}
\usepackage{hyperref}
\numberwithin{equation}{section}
\newtheorem{rules}{Rule}[section]

\newcommand{\be}{\begin{equation}}
\newcommand{\ee}{\end{equation}}
\newcommand{\ndt}{\noindent}
\def\bea{\begin{eqnarray}}
\def\eea{\end{eqnarray}}
\def\beas{\begin{eqnarray*}}
\def\eeas{\end{eqnarray*}}
\def\sla{\raise.15ex\hbox{$/$}\kern-.57em}

\def\parm{{\partial}_{-}}

\newcommand\fr[1]{\frac{1}{#1}}

\newcommand {\nn} {\nonumber}

\renewcommand{\th}{\theta}
\newcommand{\bth}{\bar\theta}

\newcommand{\dbfour}{\langle \bar d^4 \rangle}

\allowdisplaybreaks[1]

\newcommand{\er}{{\rm e}}
\newcommand{\dr}{{\rm d}}
\newcommand{\Tr}{{\rm Tr}}

\newcommand{\del}{\partial}
\newcommand{\ba}{\begin{eqnarray}}
\newcommand{\ea}{\end{eqnarray}}
\newcommand{\bdm}{\begin{displaymath}}
\newcommand{\edm}{\end{displaymath}}
\newcommand{\ra}{\rangle}
\newcommand{\la}{\langle}

\def\S{\Sigma}

\def\a{\alpha}
\def\g{\gamma}
\def\s{\sigma}

\def\veps{\varepsilon}

\def\adot{{\dot\alpha}}

\def\d{\delta}
\def\D{\Delta}
\def\v{\varphi}

\newcommand{\half}{\frac{1}{2}}

\newcommand{\ie}{{\it i.e.\ }}

\newcommand{\wtilde}{\widetilde}

\newcommand{\calG}{{\mathcal G}}

\newcommand{\calK}{{\mathcal K}}

\newcommand{\calN}{{\mathcal N}}
\newcommand{\calO}{{\mathcal O}}
\newcommand{\calP}{{\mathcal P}}
\newcommand{\calQ}{{\mathcal Q}}

\newcommand{\calS}{{\mathcal S}}

\DeclareMathAlphabet{\mathpzc}{OT1}{pzc}{m}{it}

\newcommand\hsp[1]{\hspace*{#1 cm}}
\newcommand\vsp[1]{\vspace*{#1 cm}}

\begin{document}

\begin{titlepage}
\begin{flushright}    
{\small DIAS-STP-12-03}
\end{flushright}

\vsp{1.5} 
\centerline{\Large {\bf {Gauge-invariant correlation
functions in light-cone superspace}}}

\vskip 1cm

\centerline{Sudarshan Ananth$^\dagger$, Stefano Kovacs$^*$ 
and Sarthak Parikh$^\dagger$}

\vskip .5cm

\centerline{{\it {$^\dagger$ Indian Institute of Science Education and
Research}}} \centerline{{\it {$\;$ Pune 411021, India}}}

\vskip 0.5cm

\centerline{\em $^*$ Dublin Institute for Advanced Studies}
\centerline{\em $\;$ 10 Burlington Road, Dublin 4, Ireland}

\vskip 1cm

\begin{abstract}
\ndt
We initiate a study of correlation functions of gauge-invariant operators in $\calN=4$ super Yang-Mills theory using the light-cone superspace formalism. Our primary aim is to develop efficient methods to compute perturbative corrections to correlation functions. This analysis also allows us to examine potential subtleties which may arise when calculating off-shell quantities in light-cone gauge. We comment on the intriguing possibility that the manifest $\calN=4$ supersymmetry in this approach may allow for a compact description of entire multiplets and their correlation functions.
\end{abstract}

\end{titlepage}

\section{Introduction}
\label{intro}

Correlation functions of gauge-invariant operators are the fundamental
observables in any conformally invariant gauge theory. In the case of
the $\calN=4$ supersymmetric Yang--Mills (SYM) theory the calculation
of such correlation functions has received significant attention in
recent years  because of the central role they play in the context of
the AdS/CFT correspondence~\cite{Madscft,GKP,Wadscft}. 

In this paper we initiate a systematic study of correlation functions
of gauge-invariant composite operators in $\calN=4$ SYM using the
formalism of light-cone superspace. The unique characteristic of this
formalism is that it provides a description, solely in terms of
physical degrees of freedom, in which the full $\calN=4$ supersymmetry
as well as the SU(4) R-symmetry are manifestly realised. This is
achieved at the expense of manifest Lorentz invariance. 

The formulation of $\calN=4$ SYM in light-cone superspace was
introduced in~\cite{BLN1,SM} and proved to be a powerful
tool~\cite{SM,BLN2} for the proof of the ultra-violet finiteness of
the theory to all orders in perturbation theory (see also~\cite{N4fin}). The formalism was
more recently adapted to deformations of the $\calN=4$ SYM theory with
less or no supersymmetry~\cite{LS,LM,SF} and used to prove the absence
of ultraviolet divergences in these models~\cite{AKS}. 

While the light-cone superspace formalism has a number of remarkable
properties, its application to the perturbative computation of
physical observables has been very limited.  The focus of the present
paper is on introducing the main features of $\calN=4$ light-cone
superspace as applied to the study of gauge-invariant correlation
functions in position space. To this end we present the one-loop
calculation of a simple four-point correlator of gauge-invariant
scalar operators belonging to the super-multiplet of the
energy-momentum tensor. Specifically, we will reproduce the known
tree-level and one-loop results for a four point function of composite
operators which are bilinear in the elementary scalars and transform
in the $\bm{20^\prime}$ representation of the SU(4) R-symmetry of
$\calN=4$ SYM. 

Conformal symmetry constrains the form of correlation functions and
supersymmetry further restricts the quantum corrections they
receive. This is especially the case in $\calN=4$ SYM, which has the
maximal amount of rigid supersymmetry in four dimensions. Therefore a
formalism with manifest $\calN=4$ supersymmetry is particularly
desirable and it is natural to expect that it will prove very
efficient for the calculation of correlation functions.  

Another benefit of manifest $\calN=4$ supersymmetry is the possibility
of describing an entire multiplet of operators in terms of a single
super-operator. Understanding how powerful the light-cone superspace
formulation really is in this respect will require further
investigation of the role played by the non-linearly realised
dynamical supersymmetries, but we will make some comments about this
point in the following.  

The study of gauge invariant correlation functions also allows us to
address general questions concerning the consistency of the light-cone
gauge -- and specifically its superspace implementation -- for the
calculation of off-shell quantities. This is an  important issue in
general and particularly so for a conformally invariant theory such as
$\calN=4$ SYM. In this case it is necessary to verify that no spurious
infra-red divergences are present in gauge-invariant observables. Such
spurious infra-red divergences could potentially be introduced by the
elimination of the non-physical degrees of freedom through the formal
solution of the equations of motion. Integrating out the unphysical
fields from the action leads to the appearance of $1/\del_-$
operators, which in turn could potentially introduce infra-red
singularities in Feynman diagrams. While the present paper is only a
first step towards a systematic understanding of these issues, we will
explicitly show that unphysical infra-red divergences are absent from
tree-level and one-loop corrections to a particular gauge-invariant
four-point correlation function thanks to non-trivial cancellations. 

This paper is organised as follows. In the next section we briefly
review the essential ingredients of the light-cone superspace
formulation of $\calN=4$ SYM and we present the superfield propagator
and Feynman rules in position space. In section \ref{comp-ops} we
discuss gauge-invariant super-operators and some aspects of their
correlation functions. Section \ref{4pt-correlator} contains the
calculation of a four point correlator at tree-level and
one-loop. Future directions and open problems are discussed in the
concluding section and various details of the calculations are
presented in several appendices.

\section{$\calN =4$ light cone superspace}
\label{N4-LCSS}

In this section we briefly review the formulation of $\calN=4$ SYM in
light-cone superspace~\cite{BLN1,SM}. We limit ourselves to  aspects
of the formalism which are relevant for the subsequent analysis and
refer the reader to the original papers and to more comprehensive
reviews presented in~\cite{AKS,AKP} for further details. 

\subsection{Light-cone superspace formulation of $\calN=4$ SYM}
\label{LCN4SYM}

The field content of the $\calN=4$ SYM theory comprises a gauge field,
$A_\mu$, four Weyl fermions, $\psi^m_\a$, and their conjugates,
$\bar\psi_{m \,\adot}$, $m=1,\ldots,4$, and six real scalars, $\v^i$,
$i=1,\ldots,6$. All fields transform in the adjoint representation of
the gauge group, which in the following will be taken to be
SU($N$). The theory possesses a SU(4) R-symmetry, hereby referred to
as SU(4)$_R$, under which the gauge field is a singlet, the fermions
transform in the $\mathbf{4}$ and $\mathbf{\bar 4}$ and the scalars in
the $\mathbf{6}$. 

The light-cone gauge description of the theory eliminates unphysical
degrees of freedom. The $A_-$ component of the gauge field is set to
zero (gauge choice) and the $A_+$ component is integrated out, leaving
two transverse components, $A$ and $\bar A$. Similarly half of the
components of the four Weyl fermions, $\psi^m_\a$, and their
conjugates, $\bar\psi_{m\,\adot}$, are integrated out after light-cone
projection, leaving four one-component fermionic fields, $\lambda^m$,
and their conjugates, $\bar\lambda_m$. The $\calN=4$ multiplet is
completed by the six real scalar fields, $\v^i$, $i=1,\ldots,6$, which
can be written as SU(4)$_R$ bi-spinors, $\v^{mn}$, $m,n=1,\ldots,4$,
satisfying the reality condition
\be
\bar\v_{mn} \equiv \left(\v^{mn}\right)^* = \half \veps_{mnpq}\v^{pq} \, .
\label{reality}
\ee
The relation between the two parametrisations of the scalars is
discussed in appendix \ref{app:conventions}, which also summarises
further details of the light-cone projection.

The $\calN=4$ light-cone superspace is constructed combining the four
bosonic coordinates $(x^+,x^-,x,\bar x)$ defined in (\ref{lc-ccord})
with the eight fermionic coordinates, $\theta^m$ and  $\bar\theta_m$,
$m=1,\ldots,4$, transforming in the $\mathbf{4}$ and $\mathbf{\bar 4}$
of SU(4)$_R$. In the following we will collectively denote the
superspace coordinates by $z=(x^+,x^-,x,\bar
x,\theta^m,\bar\theta_m)$. The full $\calN=4$ supersymmetry is
manifest, with half of the supercharges (denoted by $q^m$ and $\bar
q_m$, referred to as kinematical) realised linearly as translations in
the fermionic coordinates and the other half (referred to as
dynamical) non-linearly realised~\cite{BLN1,BBB,abkr}.

A central role in the light-cone superspace formulation of $\calN=4$
SYM is played by the chiral derivatives, $d^m$ and ${\bar d}_m$,
defined as 
\be
d^m = -\frac{\partial}{\partial\bar\theta_m} 
+\frac{i}{\sqrt{2}} \theta^m\parm \,, \qquad 
\bar d_m = \frac{\partial}{\partial\theta^m} 
-\frac{i}{\sqrt{2}} \bar\theta_m\parm \, , \quad m=1,\ldots,4 \, .
\label{chiralder}
\ee
They obey
\be
\{d^m,{\bar d}_n\} = i\sqrt{2}\, \d^m_n \,\del_- \, 
\label{dsusyalg}
\ee
and anti-commute with the supercharges $q^m$ and $\bar q_m$.

The component fields in the light-cone $\calN=4$ multiplet can be
packaged into a single complex superfield,
$\Phi(x,\theta,\bar\theta)$. This superfield is a SU(4)$_R$ singlet
defined by the constraints \cite{BLN1,BLN2}
\be
d^m \Phi(x,\theta,\bar\theta) = 0 \, , \qquad 
{\bar d}_m {\bar d}_n \Phi(x,\theta,\bar\theta) = \half \veps_{mnpq}
d^p d^q \bar\Phi(x,\theta,\bar\theta) \, ,
\label{origconstraints}
\ee
where $\bar\Phi = \Phi^*$ satisfies ${\bar d}_m
\bar\Phi(x,\theta,\bar\theta) =0$. The unique solution to these
constraints is a superfield with the following component 
expansion~\cite{BLN1}
\bea
\Phi\,(x,\theta,\bar\theta)&\!\!=\!\!&-\frac{1}{\parm}A(y)
-\frac{i}{\parm}\theta^m{\bar \lambda}_m(y)
+\frac{i}{\sqrt 2}\,\theta^m\theta^n{\bar \v}_{mn}(y)\nn \\
&&+\frac{\sqrt 2}{6}\theta^m\theta^n\theta^p\veps_{mnpq}\lambda^q(y) 
-\frac{1}{12}\,\theta^m\theta^n\theta^p\theta^q
\veps_{mnpq}\parm{\bar A}(y) \, ,
\label{N4supfield-1}
\eea
where the chiral coordinate 
\be
y=(x^+,y^-=x^--\frac{i}{\sqrt{2}}\theta^m\bar\theta_m,x,\bar x) 
\label{chirvar}
\ee
and the right hand side is understood to be a power expansion about
$x^-$. 

The superfields $\Phi$ and $\bar\Phi$, just like the component fields
in the $\calN=4$ multiplet, transform in the adjoint representation of
the gauge group SU($N$). They can therefore be represented as
matrices, 
\be
\Phi(x,\th,\bth) = \Phi^a(x,\th,\bth) T^a \, , \qquad 
\bar\Phi(x,\th,\bth) = \bar\Phi^a(x,\th,\bth) T^a \, , 
\ee
where $T^a$, $a=1,\ldots,N^2-1$, are generators of the fundamental
representation of SU($N$), satisfying 
\be
\Tr\left(T^aT^b\right) = \half \,\d^{ab} \, .
\label{trace-fund}
\ee
The second constraint relation in (\ref{origconstraints}) can be used
to express the conjugate superfield, $\bar\Phi(x,\th,\bth)$, in terms
of $\Phi(x,\th,\bth)$ as
\be
\bar\Phi(x,\th,\bth) = \fr{48} \frac{\dbfour}{\del_-^2} \Phi(x,\th,\bth) \, ,
\label{Phibar}
\ee
where  $\dbfour = \veps^{mnpq}\bar d_m\bar
d_n\bar d_p\bar d_q$. 

Using (\ref{Phibar}) the $\calN=4$ action in light-cone superspace can
be written purely in terms of the superfield $\Phi$ as
\ba
\calS &\!\!=\!\!& \int \dr^4x\,\dr^4\th\,\dr^4\bth \, \left\{ 
\half\Phi^a \!\left(-3\frac{\dbfour\Box}{\del_-^4}\right)\! \Phi_a 
\right. \nn \\
&& \!\!- 2gf^{abc}\left[\left(\frac{\dbfour}{\del_-^3}\Phi_a\right)\!\Phi_b
\bar\del\Phi_c 
+ \fr{48}\left(\fr{\del_-}\Phi_a\right)\!\left(
\frac{\dbfour}{\del_-^2}\Phi_b\right)\!\del\!\left(\frac{\dbfour}{\del_-^2}
\Phi_c\right)\right]  
\label{N4action-2}  \\
&&\!\!\left. -\frac{g^2}{32} f^{eab}f^{ecd}\left[\fr{\del_-}\!
\left(\Phi_a\del_-\Phi_b\right)\fr{\del_-}\!\left(\frac{\dbfour}{\del_-^2}
\Phi_c\right)\!\left(\frac{\dbfour}{\del_-} \Phi_d\right)  
+\half\Phi_a\!\left(\frac{\dbfour}{\del_-^2}
\Phi_b\right)\!\Phi_c\!\left(\frac{\dbfour}{\del_-^2}
\Phi_d\right)  \right]\right\} , \nn
\ea
where $a,b,c,d,e,\ldots=1,\ldots,N^2-1$ denote colour indices. 

Here and in the following it is understood that we use the
prescription of \cite{SM} for the $\frac{1}{\del_-}$ operator. We will
comment on potential subtleties associated with the presence of the
$\fr{\del_-}$ factors in the discussion  section. 

\subsection{Perturbative calculations in position space}
\label{pert-th-generalities}

Gauge-invariant correlation functions in a conformal field theory are
most naturally studied in position space rather than momentum space.
We now discuss some general aspects of perturbative calculations in
configuration space using the formalism of light-cone superspace. We
present the form of the superfield propagator and summarise the
Feynman rules in position space. More details are provided in the
appendices. 

The superfield propagator in position space can be obtained inverting
the kinetic operator in (\ref{N4action-2}). We start by defining the
generating functional, $Z[J]$, which gives rise to the Green functions
of the $\calN=4$ superfield upon functional differentiation with
respect to the chiral sources,  $J(x,\th,\bth)$. The Euclidean
generating functional, with the coupling to external sources suitable
for chiral superfields, is
\be
Z[J] = \frac{\displaystyle\int[\dr\Phi]\,\exp\!\left(-\calS[\Phi]+
\int\dr^{12}z\,\Phi^a(z) \frac{\la \bar d^4 \ra}{4\del_-^4} J_a(z)\right)}
{\displaystyle\int[\dr\Phi]\,\exp\left(-\calS[\Phi]\right)} \, , 
\label{gen-functnl-1}
\ee
where $z=(x,\th,\bth)$ and
$\dr^{12}z=\dr^4x\,\dr^4\th\,\dr^4\bth$. Because of the chirality
constraint that both $\Phi$ and the sources must satisfy, the rules
for functional differentiation in superspace involve subtleties which
are addressed in appendix \ref{app:prop1}. 

In order to construct the super-propagator it is sufficient to focus
on the free theory generating functional, $Z_0[J]$, obtained replacing
$\calS[\Phi]$ in (\ref{gen-functnl-1}) by the free action
$\calS_0[\Phi]$. We can write this generating functional as 
\be
Z_0[J] =  \frac{\displaystyle\int[\dr\Phi]\,\er^{-\half 
\left(\!\Phi^a,\calK_a^{\;b}\,\Phi_b\!\right) + 
\left(\!\Phi^a,\frac{\la \bar d^4 \ra}{4\del_-^4} J_a\!\right)}}
{\displaystyle\int[\dr\Phi]\,\er^{-\calS[\Phi]}} \, ,
\label{free-gen-functnl-1}
\ee
where the inner products in the exponent are
\be
-\half \left(\!\Phi^a,\calK_a^{\;b}\,\Phi_b\!\right) + 
\left(\!\Phi^a,\frac{\la \bar d^4 \ra}{4\del_-^4} J_a\!\right)
= -\half \int \dr^{12}z \,\Phi^a(z)  \left(\calK_a^{\;b}\,\Phi_b\right)\!(z)  
+ \int\dr^{12}z\,\Phi^a(z) \frac{\la \bar d^4 \ra}{4\del_-^4} J_a(z) \, .
\ee
Computing the Gaussian integral (\ref{free-gen-functnl-1}) we get
\be
Z_0[J] = \exp\left\{\half\left(\wtilde J^a,[\calK^{-1}]_a^{\;b}\,\wtilde J_b
\right) \right\} = 
\exp\left\{\half\int\dr^{12}z\,\dr^{12}z' \: \wtilde J^a(z) 
[\Delta(z-z')]_a^{\;b}\wtilde J_b(z')\right\}
\label{free-gen-functnl-2}
\ee
where 
\be
\wtilde J^a(z) = \frac{\la \bar d^4 \ra}{4\del_-^4} J^a(z) 
\ee
and the kernel, $\Delta(z-z^\prime)$, of the inverse kinetic operator,
$\calK^{-1}$, is the superfield propagator. The explicit form of
$\Delta(z-z^\prime)$ is 
\be
\Delta^a_b(z-z') = -\frac{2}{(4!)^3}
\frac{\d^a_b}{(2\pi)^2} \fr{(x-x')^2} \la d^4 \ra
\d^{(4)}(\th-\th')\d^{(4)}(\bth-\bth') \, .
\label{super-prop}
\ee 
This result is derived in detail in appendix \ref{app:prop1}. In
appendix \ref{app:prop2} we show that (\ref{super-prop}) gives rise to
the correct propagators for the component fields.

Notice that the superfield propagator in position space has
essentially the same form as in momentum space~\cite{BLN2}, apart from
the obvious replacement of $1/k^2$ by $1/(x-x^\prime)^2$.  As a
consequence the basic manipulations employed in the calculation of
position space super Feynman diagrams are also the same used in
momentum space. This represents a distinct feature compared to
covariant superspace formalisms, where there are more significant
differences between position and momentum space formulations.

The superfield interaction vertices in configuration space can be
immediately read off from the superspace action
(\ref{N4action-2}). They involve a combination of chiral and
space-time derivatives and $1/\del_-$ operators acting on the various
legs as well as group theory factors. The two cubic vertices are 
\be
\int \dr^{12}z \, (-2g)f^{abc}\left(\frac{\dbfour}{\del_-^3}\Phi_a\right)
\!\Phi_b \bar\del\Phi_c  ~~ \longrightarrow ~~ (-2g)f^{abc} 
~\raisebox{-30pt}{\includegraphics[width=4cm]{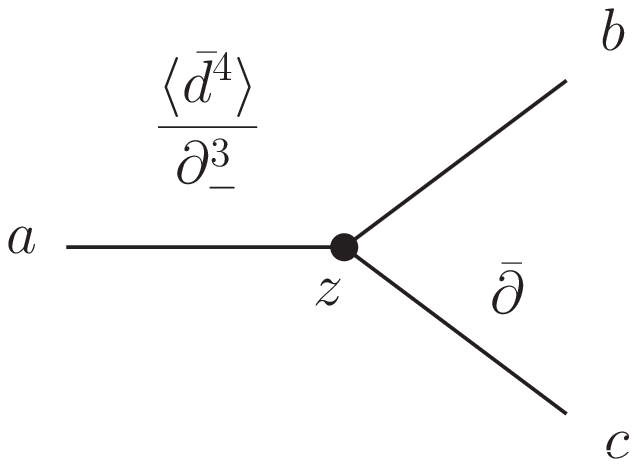}}
\label{3-vertex_1}
\ee
and 
\ba
&& \int \dr^{12}z \, \left(-\frac{g}{24}\right)f^{abc}\left(\fr{\del_-}
\Phi_a\right)\!\left(\frac{\dbfour}{\del_-^2}\Phi_b\right)\!\del\!\left(
\frac{\dbfour}{\del_-^2} \Phi_c\right) \nn \\
&& \hsp{4} \longrightarrow ~~ 
\left(-\frac{g}{24}\right)f^{abc} ~\raisebox{-38pt}{
\includegraphics[width=4cm]{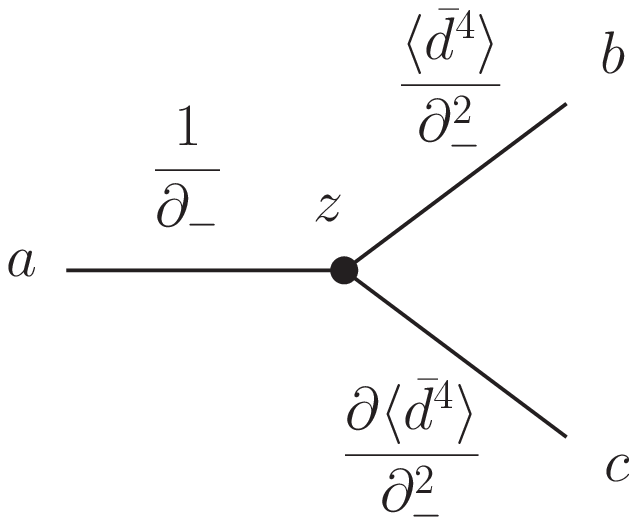}}
\label{3-vertex_2}
\ea
Here we use a black dot to denote interaction vertices, which are
integrated over the whole superspace, $z=(x,\th,\bth)$, reflecting the
fact that all intermediate steps in the calculations are manifestly
$\calN=4$ supersymmetric. Notice that the two vertices
(\ref{3-vertex_1}) and (\ref{3-vertex_2}) are complex conjugates of
each other, although this is not apparent after elimination of the
$\bar\Phi$ superfield. In the following we will refer to
(\ref{3-vertex_1})  and (\ref{3-vertex_2}) as Vertex 3-I and Vertex
3-II respectively. 

The two quartic vertices are
\ba
&& \int \dr^{12}z\,\left(-\frac{g^2}{32}\right) f^{eab}f^{ecd}
\left[\fr{\del_-}\!\left(\Phi_a\del_-\Phi_b\right)\fr{\del_-}
\!\left(\frac{\dbfour}{\del_-^2}\Phi_c\,
\frac{\dbfour}{\del_-} \Phi_d\right) \right] \nn \\
&& \hsp{5} \longrightarrow ~~ \left(-\frac{g^2}{32}\right) 
f^{eab}f^{ecd} ~ \raisebox{-44pt}{
\includegraphics[width=4cm]{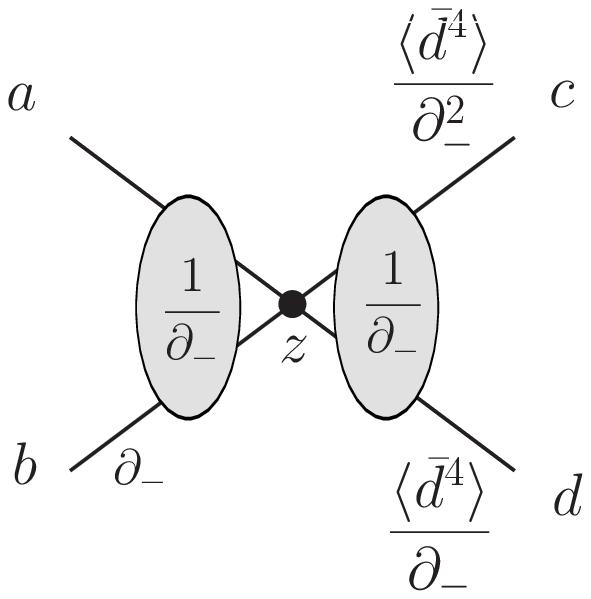}}
\label{4-vertex_1}
\ea
and
\ba
&& \int\dr^{12}z\, \left(-\frac{g^2}{64}\right) f^{eab}f^{ecd}
\left[\Phi_a\!\left(\frac{\dbfour}{\del_-^2}
\Phi_b\right)\!\Phi_c\!\left(\frac{\dbfour}{\del_-^2}
\Phi_d\right)  \right] \nn \\
&& \hsp{4} \longrightarrow ~~ \left(-\frac{g^2}{64}\right) 
f^{eab}f^{ecd} ~ \raisebox{-44pt}{
\includegraphics[width=4cm]{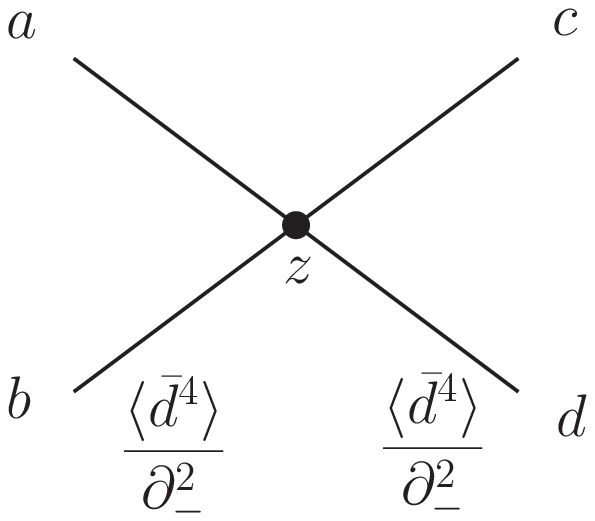}}
\label{4-vertex_2}
\ea
In the vertex (\ref{4-vertex_1}) the two $1/\del_-$ operators in the
shaded ovals act on both the adjacent legs. We will refer to
(\ref{4-vertex_1}) as Vertex 4-I and  to (\ref{4-vertex_2}) as Vertex
4-II.

Super Feynman diagrams constructed from these vertices and the
propagator (\ref{super-prop}) contain space-time derivatives ($\del$,
$\bar\del$ and $\del_-$, but not $\del_+$) as well as chiral
derivatives $d^m$ and $\bar d_m$ defined in (\ref{chiralder}). All
these derivatives can be integrated by parts in superspace
integrals. They can also be transferred from one end point to the
other of the super-propagator they act on, $\D(z-z^\prime)$,  using
the fact that the latter is only a function of the difference
$(z-z^\prime)$. Moreover, the $1/\del_-$ operators can effectively be
``integrated by parts'' as explained in (\ref{integrbyparts}). 

The general strategy for the evaluation of position space Feynman
diagrams  is similar to that used in other superspace
formulations. The first step consists in computing Grassmann
integrals, utilising the fermionic $\d$-functions in the
super-propagator. For this purpose one needs to free up one internal
line of all the chiral derivatives, using repeated integrations by
parts, and then use the relation (\ref{GRS-1}) in appendix
\ref{app:superspace-identities}. 

Once the fermionic integrals at each interaction vertex have been
computed, the external super-operators are projected onto specific
components, thus  drastically reducing the number of non-zero
contributions. 

At this point the resulting bosonic integrals can be directly compared
to the corresponding expressions obtained using Lorentz covariant
formulations. In section \ref{4pt-correlator} we illustrate these
steps in the case of a simple four-point function and we show how the
light-cone superspace analysis reproduces the known covariant results
prior to the evaluation of the final bosonic integrals.

\section{Composite operators and correlation functions}
\label{comp-ops}

Gauge-invariant operators in the $\calN=4$ SYM theory may be
classified according to their transformation properties under the
PSU(2,2$|$4) superconformal symmetry group. They can be divided into
two classes, protected operators belonging to short BPS multiplets of
the superconformal group and unprotected ones belonging to generic
long representations of PSU(2,2$|$4). BPS operators have been
classified and are characterised by shortening conditions expressed as
relations among their PSU(2,2$|$4) quantum
numbers~\cite{shortening}. Their correlation functions have special
properties and satisfy certain non-renormalisation theorems. 

In this paper we will only consider examples of correlators of
protected operators. Moreover we will confine ourselves to operators
constructed from the elementary scalars in the $\calN=4$ multiplet,
$\v^{mn}$. This ensures that the explicit form of the operators remain
the same (in light-cone gauge) as in Lorentz covariant
formulations. The simplest such operators are scalars of dimension 2
belonging  to the super-multiplet of the energy-momentum tensor, which
is a short 1/2 BPS multiplet~\footnote{It is in fact an extra-short
multiplet, with half as many components as an ordinary 1/2 BPS
multiplet.}. They transform in the representation $\bf{20^\prime}$ of 
the SU(4)$_R$ R-symmetry group and, in terms of the $\v^{mn}$ 
representation for the elementary scalars, they take the
form
\ba
Q^{[mn][pq]} &\!\!=\!\!& \Tr\left(\v^{mn}\v^{pq}\right) 
- \fr{12} \veps^{mnpq}\, \Tr\left(\bar\v_{rs}\v^{rs}\right) \nn \\
&\!\!=\!\!& \fr{3} \: \Tr\left( 2\v^{mn}\v^{pq} + \v^{mp}\v^{nq} - 
\v^{mq}\v^{np} \right) \, .
\label{d2-prot-op-2}
\ea
We can express the same operators in terms of the representation
$\v^i$ of the scalars as SU(4)$_R$ vectors as
\be
Q^{ij} = \Tr\left(\v^i\v^j\right) - \fr{6}\d^{ij}\,
\Tr\left(\v^k\v^k\right) \, .
\label{d2-prot-op-1}
\ee
The equivalence of the two forms (\ref{d2-prot-op-2}) and
(\ref{d2-prot-op-1}) can be verified using the identity
(\ref{so6su4}). 

In order to describe the operators
(\ref{d2-prot-op-2})-(\ref{d2-prot-op-1}) in light-cone superspace we
introduce composite superfield operators which contain them in their
component expansion. For this purpose it is convenient to work with
the form (\ref{d2-prot-op-1}) which, using (\ref{so6su4}) we can
rewrite as
\be
Q^{ij} = \fr{8}\left(\s^{i\,pq}\s^{j\,rs}-\fr{3}\d^{ij}
\veps^{pqrs}\right)\Tr\left(\bar\v_{pq}\bar\v_{rs}\right) \, .
\label{d2-prot-op-3}
\ee
From the form of the $\calN=4$ superfield (\ref{N4supfield-1}) and the
definition (\ref{chiralder}) of the chiral derivatives, $\bar d_m$, it
is easy to verify that the scalar field $\bar\v_{mn}(x)$ in the
expansion of $\Phi(z)$ can be isolated as follows
\be
\bar\v_{mn}(x) = \frac{i}{\sqrt{2}} \left.\left[\bar d_m \bar d_n 
\Phi(x,\th,\bth) \right]\right|_{\th=\bth=0} \, .
\label{scal-comp}
\ee
We can then define the super-operator
\be
\calQ^{ij}(z) = -\fr{16} \left( \s^{i\,pq}\s^{j\,rs}-\fr{3}\d^{ij}
\veps^{pqrs}\right) \Tr\left[\left(\bar d_p\bar d_q \Phi(z)\right)
\left(\bar d_r \bar d_s \Phi(z)\right)\right] \, , 
\label{superQ}
\ee
which contains (\ref{d2-prot-op-3}) as its $\th=\bth=0$ component,
\be
Q^{ij}(x) = \left.\left[ \calQ^{ij}(z)\right]\right|_{\th=\bth=0} \, .
\label{Q-calQ-rel}
\ee
The only other operator of bare dimension 2 in the $\calN=4$ theory is
an unprotected one, the superconformal primary operator, $K(x)$,
belonging to the long Konishi multiplet~\cite{konishi,more-konishi,bkrs-k}. $K(x)$
is a SU(4)$_R$ singlet and takes the form
\be
K = \Tr(\v^i\v^i) = \fr{4}\veps^{mnpq}\Tr
\left(\bar\v_{mn}\bar\v_{pq}\right) \, .
\label{d2-konishi}
\ee
Using (\ref{scal-comp}) we can construct a super-operator containing
$K(x)$ as $\th=\bth=0$ component. We define 
\be
\calK(z) = -\fr{8}\veps^{mnpq} \Tr\left[\left(\bar d_p\bar d_q \Phi(z)
\right)\left(\bar d_r \bar d_s \Phi(z)\right)\right] \, ,
\label{superK}
\ee
so that
\be
K(x) = \left.\left[ \calK(z)\right]\right|_{\th=\bth=0} \, .
\label{K-calK-rel}
\ee

The conformal symmetry of $\calN=4$ SYM constrains the
correlation functions of gauge-invariant primary operators~\cite{CFT}. The
space-time dependence of two-point functions of primary operators,
$\calO_k(x)$, is completely fixed by conformal invariance, 
\be
\la \calO_h(x) \calO_k(y) \ra = \frac{c_{hk}^{(2)}}{(x-y)^{2\D_k}} \, ,
\label{gen-2pt-funct}
\ee
where $\D_k$ is the common scaling dimension of $\calO_k$ and
$\calO_h$. In general $\D_k$ (as well as the coefficient
$c^{(2)}_{kh}$) receives quantum corrections,
\be
\D_k = \D_k^{(0)} + \g_k(g) \, ,
\label{anom-dim}
\ee
where $\g_k(g)$ is the anomalous dimension~\footnote{This is a slight
oversimplification. An accurate definition of the anomalous dimension
involves the diagonalisation of the set of two-point functions
(\ref{gen-2pt-funct}), \ie the resolution of the mixing among
operators of the same bare dimension.}. The study of the spectrum of anomalous dimensions of composite operators in $\calN=4$ has been a major focus of activity in recent years, in particular in connection with the emergence of integrability properties~\cite{integra}.

Three-point functions of primary operators are also constrained in
their form, 
\be
\la \calO_h(x)\calO_k(y)\calO_l(z)\ra = \frac{c^{(3)}_{hkl}}
{(x-y)^{\D_h+\D_k-\D_l}(x-z)^{\D_h+\D_l-\D_k}(y-z)^{\D_k+\D_l-\D_h}}
\, , 
\label{gen-3pt-funct}
\ee
where the numerator, $c^{(3)}_{hkl}$, is related to the Operator
Product Expansion (OPE) coefficients for the three operators and in
general receives quantum corrections. 

As mentioned earlier, BPS operators in $\calN=4$ SYM have special
non-renormalisation properties. For such operators the BPS condition
implies a relation between their SU(4)$_R$ quantum numbers and their
scaling dimensions which, as a consequence, do not receive quantum
corrections. This fact in turn is related to the absence of quantum
corrections to the two-point correlation functions of BPS
operators~\cite{non-renorm-2}.  Similarly three-point functions of BPS
operators in $\calN=4$ SYM are tree-level exact and this is related to
the absence of quantum corrections  to the OPE coefficients among
triplets of protected operators~\cite{non-renorm-3}.  Four- and higher-point functions of
protected operators do receive non-trivial quantum corrections both in
perturbation theory~\cite{bkrs-1,bkrs-2,pert-higher-pt} and from
instantons~\cite{instantons}, but they are ultra-violet finite. 

In general all correlation functions of non-protected operators
receive quantum corrections, including two- and three-point
functions. Even in a finite and exactly conformally invariant theory
such as $\calN=4$ SYM, these corrections are accompanied by
ultra-violet infinities~\cite{bkrs-k}. These divergences are an artefact of the
perturbative expansion and can be reabsorbed into the renormalisation
of scaling dimensions and OPE coefficients. However, their presence
implies the need to introduce a regularisation scheme even in
$\calN=4$ SYM and thus leads to additional subtleties. 

In the present paper we consider only correlators of protected
operators, focussing on a four-point function of the $\calQ^{ij}$
defined in (\ref{superQ}). In the following section we present the
tree-level and one-loop calculations for this four-point function and
in deriving our results we will assume the non-renormalisation of two-
and three-point functions discussed above. Although the
non-renormalisation results were obtained in covariant gauges, the
absence of corrections to scaling dimensions and OPE coefficients of
BPS operators is a gauge-invariant result and thus remains valid when
working in light-cone superspace. Moreover, the fact that these scalar
operators have the same explicit form as in covariant gauges should
also ensure the absence of non-physical quantum corrections, such as
those associated with a wave function renormalisation, to their two-
and three-point functions. 

Loop corrections to two- and three-point functions of protected
operators  involve divergent integrals, which arise in conjunction
with vanishing coefficients. Therefore an explicit proof of the
non-renormalisation of these correlators requires the introduction of
a suitable regularisation. We intend to return to a more detailed
analysis of this matter in a future publication, where we will address
the issues associated with the regularisation of light-cone superspace
calculations in the context of the study of more general non-protected
correlation functions.

One of the benefits of superspace formulations of supersymmetric gauge
theories is the possibility of providing a compact description of
entire multiplets in terms of superfields. In this respect the
light-cone superspace description of $\calN=4$ SYM is particularly
interesting as it is the only formulation of the theory in which the
full $\calN=4$ supersymmetry is manifest. Working with super-operators
such as (\ref{superQ}) and (\ref{superK}) should make it possible to
extract all correlation functions of operators in the same
supersymmetry multiplet from a single super-correlator. It will be
interesting to study other components in the $\th$-expansion of the
super-correlation function considered in the next section. These
should contain information about correlation functions of the
super-partners of the $Q^{ij}$'s. 

The multiplet starting with the superconformal primary operator
(\ref{d2-prot-op-2})-(\ref{d2-prot-op-1}) contains the conserved
currents associated with the PSU(2,2$|$4) superconformal symmetry of
$\calN=4$ SYM, \ie  the energy-momentum tensor, the supersymmetry and
R-symmetry currents, as well as other bosonic and fermionic operators
for a total of 128+128 components. All these operators are written as
SU($N$) traces of products of two, three or four elementary fields in
the $\calN=4$ multiplet. 

Although in this paper we are concerned only with correlation
functions of the superconformal primaries
(\ref{d2-prot-op-2})-(\ref{d2-prot-op-1}), it is natural to speculate
that the light-cone superspace formalism will permit a description of
the entire energy-momentum tensor multiplet using a single composite
superfield. This will require the addition of terms cubic and quartic
in the superfield $\Phi$ to the super-operator (\ref{superQ}). These
additional terms should not modify the $\th=\bth=0$ component, while
producing the correct cubic and quartic terms in the remaining
operators. The exact form of these additional terms in the
super-operator should be determined by the entire $\calN=4$
superalgebra, including the non-linearly realised dynamical
generators. The possibility of constructing such a composite
superfield operator is intriguing and we hope to investigate it
further.

\section{A simple four-point correlation function}
\label{4pt-correlator}

The study of four-point correlation functions of protected operators
in $\calN=4$ SYM provides an ideal testing ground for the application
of light-cone superspace techniques to the calculation of off-shell
observables. As mentioned in the previous section, four-point
functions of BPS operators are less constrained by (super)conformal
invariance than two- and three-point functions. They receive quantum
corrections, but are free of both infra-red and ultra-violet
divergences for generic positions of the operator insertions. 

In the case of four-point functions of $\calN=4$ primary operators the
dependence on the external points is not fixed by the symmetries of
the theory. Quantum corrections to these correlators can be
reorganised into functions, $F_4(r,s;g)$, of the coupling constant and
two conformally invariant cross ratios, which can be chosen as
\be
r = \frac{x_{12}^2x_{34}^2}{x_{13}^2x_{24}^2} \, , \qquad
s = \frac{x_{14}^2x_{23}^2}{x_{13}^2x_{24}^2} \, , 
\label{cross-ratios}
\ee
where $x_{ij}^2 = (x_i-x_j)^2$. The functions $F_4(r,s;g)$ are
unrestricted by the PSU(2,2$|$4) global symmetry group and, in
general, receive an infinite series of perturbative and instanton
corrections.  It is also worth recalling that four-point correlation
functions of protected operators contain information on anomalous
dimensions and OPE coefficients of non-protected operators in the
$\calN=4$ SYM spectrum, which can be extracted from singularities
arising in short distance limits, $x^2_{ij}\to 0$, for pairs of
external points~\cite{OPE}.

We consider four-point correlation functions of the operators $Q^{ij}$
given in (\ref{d2-prot-op-3}),
\be
G^{(Q)}_4(x_1,\dots,x_4) = \la Q^{i_1j_1}(x_1)\,Q^{i_2j_2}(x_2)\,
Q^{i_3j_3}(x_3)\,Q^{i_4j_4}(x_4) \ra \, ,
\label{G4Q-def}
\ee
which can be obtained from the correlation functions of the
corresponding super-operators, $\calQ^{ij}$, defined as
\be
\calG^{(Q)}_4(z_1,\dots,z_4) = \la \calQ^{i_1j_1}(z_1)\,
\calQ^{i_2j_2}(z_2)\, \calQ^{i_3j_3}(z_3)\,\calQ^{i_4j_4}(z_4) \ra \, ,
\label{superG4Q-def}
\ee
by setting to zero the external fermionic coordinates, 
\be
G^{(Q)}(x_1,\ldots,x_4) = \left. \calG_4^{(Q)}(z_1,\ldots,z_4)
\right|_{\th^{(\a)m} = \bth^{(\a)}_m = 0} \, , \qquad 
\forall\; \a =1,\ldots,4\, ,
\;m=1,\ldots,4 \, ,
\label{G4Q-superG4Q}
\ee
where the index $\a$ labels the external points.

The super-operators $\calQ^{ij}$ defined in (\ref{superQ}) and their
lowest components (\ref{d2-prot-op-3}) transform in the representation
$\bm{20^\prime}$ of SU(4)$_R$. As a consequence there are in principle
six independent four-point functions of the type in
(\ref{superG4Q-def}), (\ref{G4Q-def}), since the singlet enters with
multiplicity 6 in the tensor product of four
$\bm{20^\prime}$'s. However, explicit perturbative and instanton
calculations indicate that there exist functional relations among any
six four-point functions that can be chosen as a basis, leaving only
two independent structures. Therefore all correlation functions in
(\ref{G4Q-def}) are determined by two independent functions,
$F_4^{(1)}(r,s;g)$ and $F_4^{(2)}(r,s;g)$, of the cross ratios
(\ref{cross-ratios}). 

In this paper we restrict our attention to a simple four-point
function in the class (\ref{G4Q-def}), which we denote by
$G_4^{(H)}(x_1,\dots,x_4)$. It corresponds to the following choice for
the flavour indices
\be
G^{(H)}_4(x_1,\dots,x_4) = \la Q^{12}(x_1)\,Q^{34}(x_2)\,
Q^{34}(x_3)\,Q^{12}(x_4) \ra \, .
\label{G4H-def}
\ee
We re-derive the known tree-level and one-loop contributions to
(\ref{G4H-def}) working in light-cone superspace. Our starting point
is thus 
\be
\calG^{(H)}_4(z_1,\dots,z_4) = \la \calQ^{12}(z_1)\,\calQ^{34}(z_2)\,
\calQ^{34}(z_3)\,\calQ^{12}(z_4) \ra \, ,
\label{superG4H-def}
\ee
which reduces to (\ref{G4H-def}) upon setting to zero the external
fermionic coordinates. 

The simplifications induced by the choice of SU(4)$_R$ indices in
(\ref{G4H-def}) will become apparent in the next subsections where we
evaluate this particular four-point function at tree-level and
one-loop. 

We start by writing (\ref{G4H-def}) using the form
(\ref{d2-prot-op-3}) for the $Q^{ij}$ operators,
\bea
&& \hsp{-.6}G_4^{(H)}(x_1,\ldots,x_4) 
= \Big\la \Tr\left[\left(\v^{1}\v^{2}\right)\!(x_1)\right] 
\Tr\left[\left(\v^{3}\v^{4}\right)\!(x_2)\right]
\Tr\left[\left(\v^{3}\v^{4}\right)\!(x_3)\right] 
\Tr\left[\left(\v^{1}\v^{2}\right)\!(x_4)\right]  \Big\ra  \nn \\
&& \hsp{-.6} = \left(\fr{8}\right)^4\sigma^{1m_1n_1}
\sigma^{2p_1q_1}\sigma^{3p_2q_2}\sigma^{4m_2n_2}
\sigma^{3p_3q_3}\sigma^{4m_3n_3}\sigma^{1m_4n_4}
\sigma^{2p_4q_4} \label{tree-level-2} \rule{0pt}{18pt} \\
&& \hsp{-.6}\times \Big\la \Tr\left[\left(\bar\v_{m_1n_1}
\bar\v_{p_1q_1}\right)\!(x_1)\right] 
\Tr\left[\left(\bar\v_{m_2n_2}\bar\v_{p_2q_2}\right)\!(x_2)\right]
\Tr\left[\left(\bar\v_{m_3n_3}\bar\v_{p_3q_3}\right)\!(x_3)\right] 
\Tr\left[\left(\bar\v_{m_4n_4}\bar\v_{p_4q_4}\right)(x_4)\right]  
\Big\ra\,. \nn \rule[-10pt]{0pt}{28pt}
\eea
The explicit form of the super-operator containing
(\ref{tree-level-2}) as its $\th=\bth=0$ component is 
\ba 
&&\hsp{-1.5}\calG^{(H)}_4(z_1,\dots,z_4) =
\fr{16}\left(\fr{8}\right)^4 \left(\!\frac{i}{\sqrt{2}}\!\right)^8
\sigma^{1m_1n_1}\sigma^{2p_1q_1}\sigma^{3p_2q_2}
\sigma^{4m_2n_2}\sigma^{3p_3q_3}\sigma^{4m_3n_3}
\sigma^{1m_4n_4}\sigma^{2p_4q_4} \nn\\
&& \times \Big\la\Big( \bar d_{m_1}^{(1)}\bar d_{n_1}^{(1)}
\Phi^a(z_1)\bar d_{p_1}^{(1)}\bar d_{q_1}^{(1)}\Phi^a(z_1)\!\Big) 
\!\Big(\bar d_{m_2}^{(2)}\bar d_{n_2}^{(2)}\Phi^b(z_2)
\bar d_{p_2}^{(2)}\bar d_{q_2}^{(2)}\Phi^b(z_2)\Big)\! 
\Big. \nn \\ \nn \\
&& \times \Big(\bar d_{m_3}^{(3)}\bar d_{n_3}^{(3)}
\Phi^c(z_3)\bar d_{p_3}^{(3)}\bar d_{q_3}^{(3)}\Phi^c(z_3)\Big)
\!\Big(\bar d_{m_4}^{(4)}\bar d_{n_4}^{(4)}\Phi^d(z_4)
\bar d_{p_4}^{(4)}\bar d_{q_4}^{(4)}\Phi^d(z_4)\Big)
\Big\ra\,.
\label{4pt}
\ea
Notice that in $G_4^{(H)}(x_1,\ldots,x_4)$ we choose all the $Q^{ij}$
operators with distinct flavour indices, so that when re-writing them
in the form (\ref{d2-prot-op-3}) the second term, which subtracts the
SU(4)$_R$ trace never appears. This leads to simplifications in the
calculation since there are fewer contractions to consider. In
addition, divergences in intermediate steps are avoided. The flavour
trace that is subtracted in (\ref{d2-prot-op-3}) is in fact
proportional to the unprotected Konishi operator (\ref{d2-konishi}),
whose four-point functions are divergent at any fixed order in
perturbation theory. These divergences would be needed, had we chosen
to insert $Q^{ij}$ operators with $i=j$, to cancel other divergent
contributions and ensure the finiteness of the resulting $G_4^{(Q)}$
four-point function.

For compactness of notation, in the following we write the
super-propagator as 
\be
\D^i_j(z_1-z_2) = \frac{k \d^i_j}{x_{12}^2} \la d^4 \ra \d^{(8)}_{12} \, ,
\label{comp-prop}
\ee
where $k=-2/(2\pi)^2(4!)^3$, $x_{12}^2=(x_1-x_2)^2$ and $\d^{(8)}_{12}
= \d^{(4)}(\th_1-\th_2)\d^{(4)}(\bth_1-\bth_2)$. 

\subsection{Tree level}

\begin{figure}[htb]
\begin{center}
\includegraphics[width=0.4\textwidth]{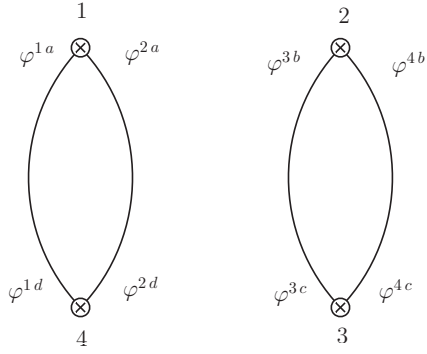}
\end{center}
\caption{Tree-level contribution to $G^{(H)}_4(x_1,\ldots,x_4)$.}
\label{fig:disc-tree}
\end{figure}

At tree level there are multiple contractions possible in
(\ref{4pt}). However, only the one shown in figure \ref{fig:disc-tree}
is non-zero. The reason why all other contractions vanish is evident
from the form of $G_4^{(H)}$ in the first line of
(\ref{tree-level-2}): all other contraction are zero because the
propagator (\ref{scalar-prop}) for the elementary scalars is diagonal
in flavour space. 

It is straightforward to obtain the same result in superspace. A free
propagator connecting scalars $\v^{a_1\,i_1}(x_1)$ and
$\v^{a_2\,i_2}(x_2)$ in two $Q$ operators gives rise to the factor
\be
\s^{i_1m_1n_1}\s^{i_2m_2n_2}\left(\bar d_{m_1}\bar d_{n_1} 
\la d^4\ra\,\frac{\d^{(8)}_{12}}{x^2_{12}}\,
\overleftarrow{\bar d_{m_2} \bar d_{n_2}}\,\d^{a_1a_2}\right) \, ,
\ee
which, upon setting to zero the $\th$ and $\bth$ coordinates at points
$z_1$ and $z_2$, reduces to 
\be
\left. \frac{\d^{a_1a_2}}{x_{12}^2}\, \s^{i_1m_1n_1}\s^{i_2m_2n_2}
\la d^4\ra \bar d_{m_1}\bar d_{n_1} 
\bar d_{m_2}\bar d_{n_2} \,\d^{(8)}_{12}
\right|_{\th_1=\bth_1=0} =(4!)^3  \,8\, \frac{\d^{a_1a_2}}{x_{12}^2} 
\,\d^{i_1i_2} \, ,
\ee
where $\sigma^{i_1m_1n_1}\sigma^{i_2m_2n_2}\veps_{m_1n_1m_2n_2} =
8\,\d^{i_1i_2}$. Thus each external $\v^i$ can only be connected
through a free propagator to a $\v^j$ with $i=j$ for a non-vanishing
contribution. Therefore at tree level the only allowed contraction in
$\calG_4^{(H)}(z_1,\ldots,z_4)$ is the one in figure
\ref{fig:disc-tree} which, using (\ref{comp-prop}), yields 
\ba
&& \hsp{-1.5} \calG^{(H)}_4(z_1,\ldots,z_4) = \fr{2^{20}}\,
\sigma^{1m_1n_1}\sigma^{2p_1q_1}\sigma^{3p_2q_2}\sigma^{4m_2n_2}
\sigma^{3p_3q_3}\sigma^{4m_3n_3}\sigma^{1m_4n_4}\sigma^{2p_4q_4} \nn\\
&&\hsp{1.25}\times \,k^4\!\left(\bar d_{m_1}\bar d_{n_1} \la d^4\ra 
\frac{\d^{(8)}_{14}}{x^2_{14}} \overleftarrow{\bar d_{n_4}
\bar d_{m_4}}\d^{ad}\right)\!\!\left(\bar d_{p_1}\bar d_{q_1} 
\la d^4\ra \frac{\d^{(8)}_{14}}{x^2_{14}} 
\overleftarrow{\bar d_{q_4}\bar d_{p_4}}\d^{ad}\right)\nn\\
&&\hsp{1.25}\times\left(\bar d_{m_2}\bar d_{n_2} \la d^4\ra 
\frac{\d^{(8)}_{23}}{x^2_{23}} \overleftarrow{\bar d_{n_3}
\bar d_{m_3}}\d^{bc}\right)\!\!\left(\bar d_{p_2}\bar d_{q_2} 
\la d^4\ra \frac{\d^{(8)}_{23}}{x^2_{23}} 
\overleftarrow{\bar d_{q_3}\bar d_{p_3}}\d^{bc}\right) \, .
\ea
Setting to zero all the external $\th^m$'s and $\bth_m$'s we get
\ba
&&  \hsp{-0.6} G^{(H)}_4(x_1,\ldots,x_4) = 
\left(\sigma^{1m_1n_1}\sigma^{1m_4n_4}\veps_{m_1n_1m_4n_4}\right)
\left(\sigma^{2p_1q_1}\sigma^{2p_4q_4}\veps_{p_1q_1p_4q_4}\right)
\left(\sigma^{3p_2q_2}\sigma^{3p_3q_3}\veps_{p_2q_2p_3q_3}\right)\nn\\
&&  \hsp{1}\times \left(\sigma^{4m_2n_2}\sigma^{4m_3n_3}
\veps_{m_2n_2m_3n_3}\right)\fr{2^{20}}\, k^4  
(N^2-1)^2 (4!)^{12}\fr{(x^2_{14})^2(x^2_{23})^2}\,,
\label{disc-tree-1}
\ea
where we used $\d^{aa}=N^2-1$. Simplifying
(\ref{disc-tree-1}) and substituting $k=-2/(2\pi)^2(4!)^3$ we get 

\vsp{0.5}

\be
\left[G_4^{(H)}(x_1,\ldots,x_4)\right]_{\rm tree} 
= \frac{\left(N^2-1\right)^2}{16 (2\pi)^8}
\fr{(x^2_{14})^2(x^2_{23})^2}\,.
\label{final-tree}
\ee

\vsp{0.1}

\subsection{One-loop}

One-loop contributions to $G_4^{(H)}(x_1,\ldots,x_4)$ are of order
$g^2$ and involve either two cubic interaction vertices or a single
quartic vertex. Moreover we can distinguish between disconnected
diagrams, which factorise into the product of tree-level and one-loop
two-point functions, and connected four-point diagrams.

\subsubsection{Factorised two-point functions} 

\begin{figure}[htb]
\begin{center}
\subfloat[]{\label{fig:disc-cubic-3}
\includegraphics[width=0.39\textwidth]{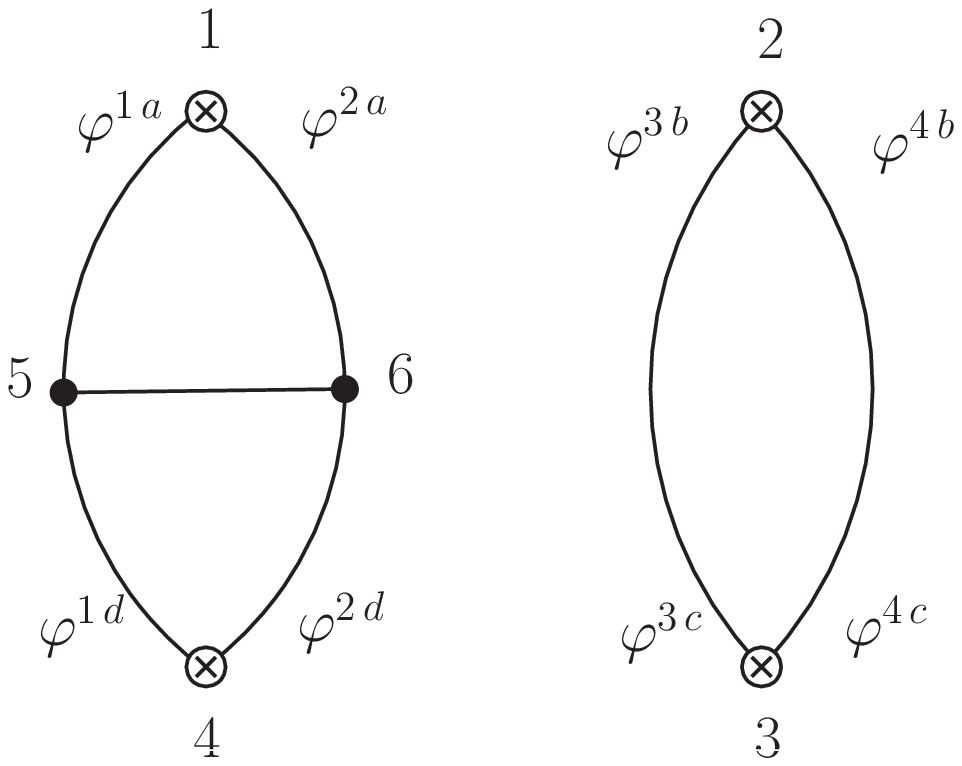} }
\qquad\subfloat[]{\label{fig:disc-cubic-4}
\includegraphics[width=0.39\textwidth]{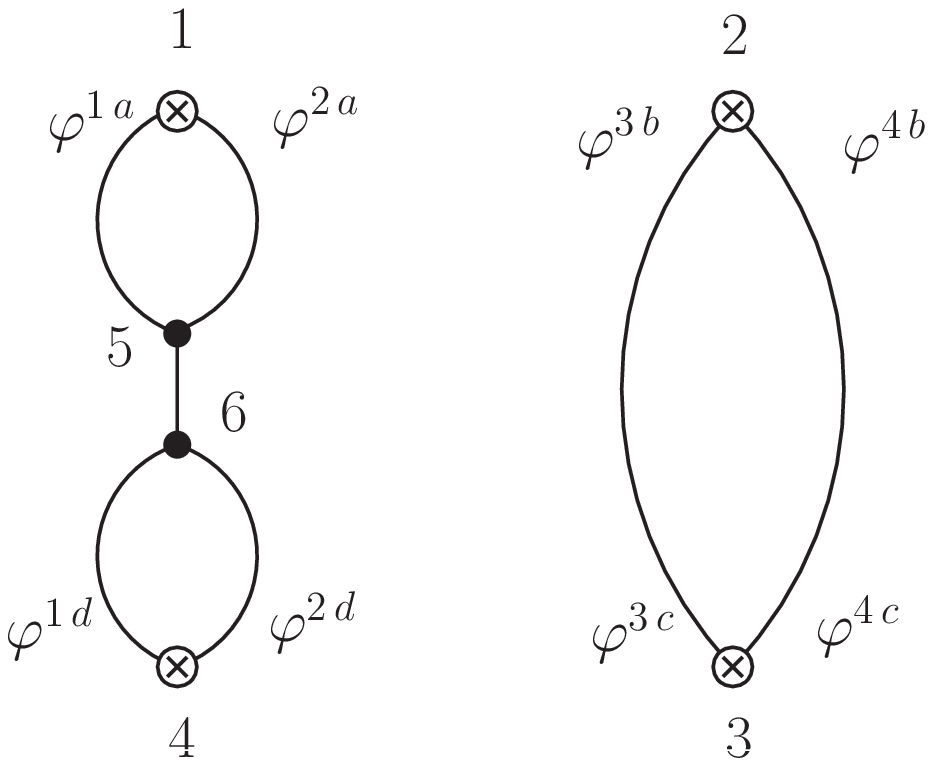} }
\vskip 0.3cm
\subfloat[]{\label{fig:disc-cubic-5}
\includegraphics[width=0.42\textwidth]{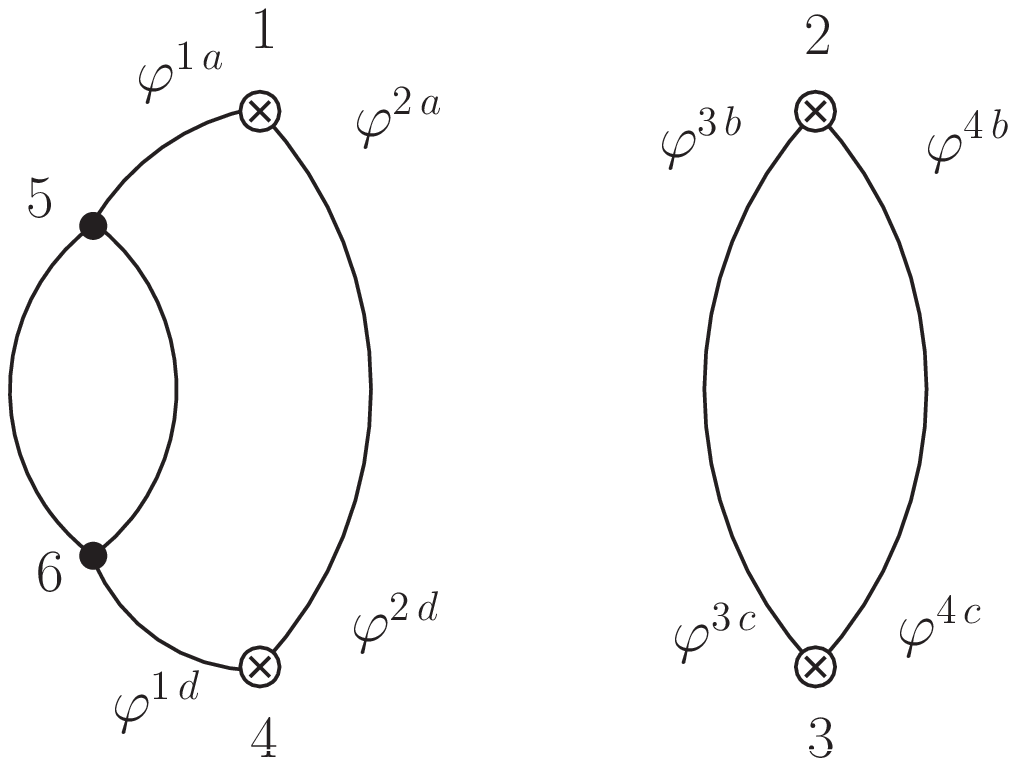} }
\qquad
\subfloat[]{\label{fig:disc-quartic-2}
\includegraphics[width=0.39\textwidth]{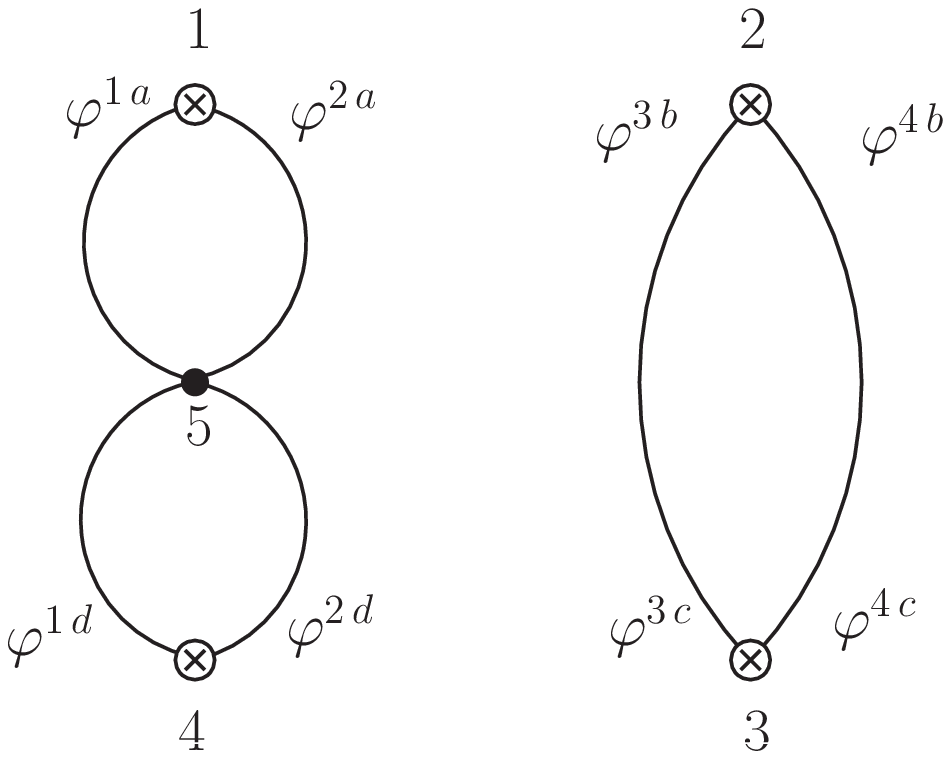}}
\end{center}
\caption{Disconnected one-loop contributions to $G_4^{(H)}(x_1,\ldots,x_4)$.}
\label{fig:disc-cubic-2pt}
\end{figure}

Figure \ref{fig:disc-cubic-2pt} depicts the disconnected one-loop
contributions to $G_4^{(H)}$. They factorise as
\be
\la \calQ^{12}(z_1)\,\calQ^{12}(z_4) \ra_{\rm 1-loop} 
\la \calQ^{34}(z_2)\,\calQ^{34}(z_3) \ra_{\rm tree} \, .
\label{factor-1}
\ee
A second set of diagrams in which the interaction vertices connect to
the external points $z_2$ and $z_3$ gives rise to a contribution of
the form
\be
\la \calQ^{12}(z_1)\,\calQ^{12}(z_4) \ra_{\rm tree} 
\la \calQ^{34}(z_2)\,\calQ^{34}(z_3) \ra_{\rm 1-loop} \, .
\label{factor-2}
\ee
Both (\ref{factor-1}) and (\ref{factor-2}) vanish thanks to the
non-renormalisation of two-point functions of protected operators.
Therefore we assume that $G_4^{(H)}(x_1,\ldots,x_4)$ receives no
contribution from the sum of all diagrams with the topologies in
figure \ref{fig:disc-cubic-2pt}.  While this assumption is justified
because the vanishing of one-loop corrections to two point functions
of BPS operators is a gauge-independent result, it would be desirable
to have an explicit proof in light-cone superspace and we intend to
revisit this issue.

\subsubsection{Connected diagrams involving two cubic vertices}
\label{connect-cubic}

\begin{figure}[htb]
\begin{center}
\subfloat[]{\label{fig:disc-cubic-1}
\includegraphics[width=0.35\textwidth]{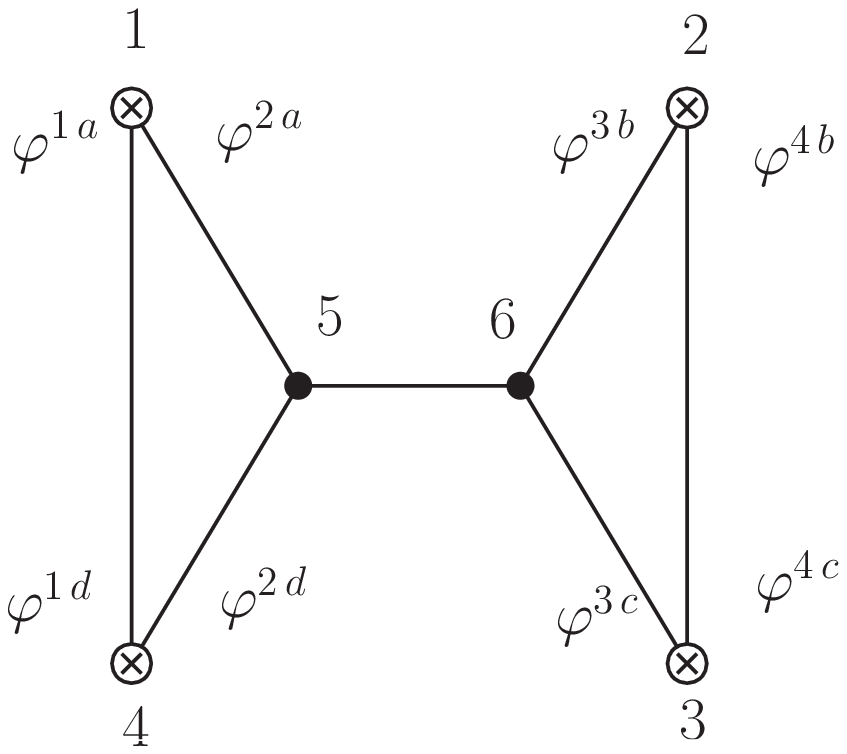} }
\qquad \subfloat[]{\label{fig:disc-cubic-2}
\includegraphics[width=0.35\textwidth]{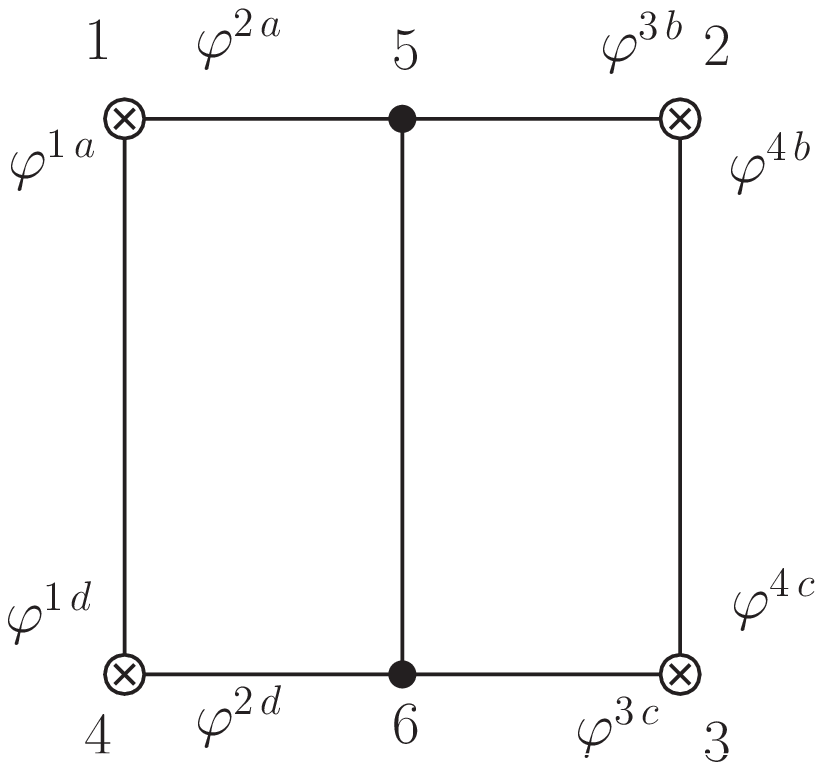} }
\end{center}
\caption{Connected one-loop contributions to
$G_4^{(H)}(x_1,\ldots,x_4)$ involving cubic vertices.}
\label{fig:disc-cubic}
\end{figure}

The next set of diagrams of order $g^2$ that we need to consider are
connected ones involving two cubic vertices. There are two distinct
types of contractions to take into account which are shown in figure
\ref{fig:disc-cubic}. 

The building blocks for these diagrams are the cubic vertices
(\ref{3-vertex_1}) and (\ref{3-vertex_2}). Analysing the combinations
of chiral derivatives in these vertices one can verify that in order
to produce a potentially non-vanishing contribution a diagram must
involve one vertex of each type. This is proven in appendix
\ref{app:rules-cubic}, where we also discuss an explicit example. A
consequence of this observation is the reality of individual
contributions to the four-point functions $G_4^{(Q)}$, as the two
cubic vertices (\ref{3-vertex_1}) and (\ref{3-vertex_2}) are complex
conjugates of each other.

The contributions from the two diagrams in figure \ref{fig:disc-cubic}
vanish individually, but for different reasons. 

The vanishing of diagrams of the type in figure \ref{fig:disc-cubic-1}
is straightforward. Since the superfield propagator is diagonal in
colour space, the free contractions between points $z_1$ and $z_4$ and
between points $z_2$ and $z_3$, combined with the traces at each
external point, force two of the indices of the totally antisymmetric
structure constants $f^{abc}$ at the interaction vertices in $z_5$ and
$z_6$ to be the same. Therefore these diagrams are identically
zero. Since this vanishing result follows from the colour structure of
the diagram, all other Wick contractions, which differ only in the
distribution of flavour indices, give a zero result as well.

\vsp{0.3}
Diagrams of the type shown in figure \ref{fig:disc-cubic-2} also
vanish, but the proof is slightly more involved, requiring
manipulations which are described in detail in appendix
\ref{app:rules-cubic}. The vanishing of contributions with this
topology follows from the observation that a contraction in which two
external fields $\v^{i_1}$ and $\v^{i_2}$ are connected to a cubic
interaction vertex gives rise to a factor of
$\sigma^{i_1mn}\sigma^{i_2pq}\veps_{mnpq}=8\d^{i_1i_2}$. The reason
for this is explained under Rule \ref{rules-cubic} in Appendix
\ref{app:rules-cubic}.

In the case of the diagram  in figure \ref{fig:disc-cubic-2} the
internal point $z_5$ ($z_6$) connects $\v^2$ with $\v^3$, which
results in a factor of $\sigma^{2mn}\sigma^{3pq}\veps_{mnpq} = 0$.
Other Wick contractions, with a different distribution of flavour
indices, vanish for the same reason.

In general since $\sigma^{i_1mn}\sigma^{i_2pq}\veps_{mnpq} = 0$ for
$i_1\neq i_2$, scalar fields $\v^{i_1}$ and  $\v^{i_2}$ with $i_1\neq
i_2$ cannot be connected through a cubic vertex.

\subsubsection{Connected diagrams involving one quartic vertex}
\label{quart-conn}

\begin{figure}[htb]
\begin{center}
\includegraphics[width=0.35\textwidth]{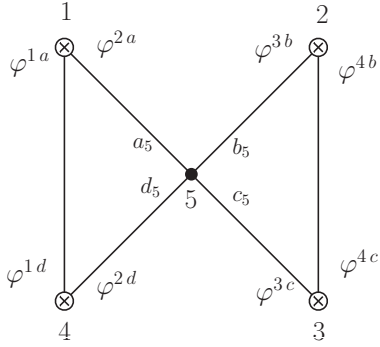}
\end{center}
\caption{Connected one-loop contributions to
$G_4^{(H)}(x_1,\ldots,x_4)$ involving a quartic vertex 
($a_5$, $b_5$, $c_5$ and $d_5$ are colour indices).}
\label{fig:disc-quartic}
\end{figure}

The last type of contribution to $G_4^{(H)}(x_1,\dots,x_4)$ at order
$g^2$ comes from diagrams involving a single quartic vertex. With our
choice of external flavours the only allowed topology is depicted in
figure \ref{fig:disc-quartic}, where the interaction vertex at point
$z_5$ can be either (\ref{4-vertex_1}) or (\ref{4-vertex_2}). The
first type of contribution, constructed using the vertex
(\ref{4-vertex_1}), vanishes. Therefore the entire one-loop correction
to $G_4^{(H)}(x_1,\dots,x_4)$ comes from diagrams of the type in
figure \ref{fig:disc-quartic}, with the quartic interaction at point
$z_5$ corresponding to Vertex 4-II (\ref{4-vertex_2}). 

We present in detail the calculation of the contraction shown in the
figure, in which the two free propagators connecting points $z_1$ and
$z_4$ and points $z_2$ and $z_3$ carry flavour 1 and 4
respectively. There are additional contributions in which the
$z_1-z_4$ line has flavour 2 and/or the $z_2-z_3$ line has flavour
3. These produce the same contribution as the diagram we analyse and
therefore simply give rise to a multiplicity factor in the final
answer. 

The vanishing of diagrams involving Vertex 4-I (\ref{4-vertex_1}) can
be understood as follows.  In this figure, colour labels $a_5$ and
$b_5$ ($c_5$ and $d_5$) cannot sit on both the interaction legs to the
left, or both the legs to the right, else the structure function at
the interaction point, $f^{ea_5b_5}f^{ec_5d_5}$ will vanish. This is
due to the external propagator connecting $x_1$ to $x_4$ ($x_2$ to
$x_3$) -- the Kronecker delta in the propagator forces the two colour
indices on the two left (right) legs of the interaction vertex to be
the same.

However, unless $a_5$ and $b_5$ ($c_5$ and $d_5$) sit on both the left
legs or both the right legs, we will run into a contraction of the
form $\sigma^{2mn}\sigma^{3pq}\veps_{mnpq}$, which is zero. The reason
why we end up with this contraction is explained under Rule
\ref{rules-quartic} in Appendix \ref{app:rules-quartic}.  Thus the
requirement that the structure functions be non-zero conflicts with
the requirement that the  $\sigma\sigma\veps$ contractions be
non-zero. Consequently Vertex 4-I does not contribute.

\vsp{0.3}
Finally we come to the calculation of the non-zero contribution from
diagrams of the topology in figure \ref{fig:disc-quartic} in which the
interaction vertex is of type 4-II. 

\begin{figure}[htb]
\begin{center}
\subfloat[]{\label{fig:disc-quart-split1}
\includegraphics[width=0.35\textwidth]{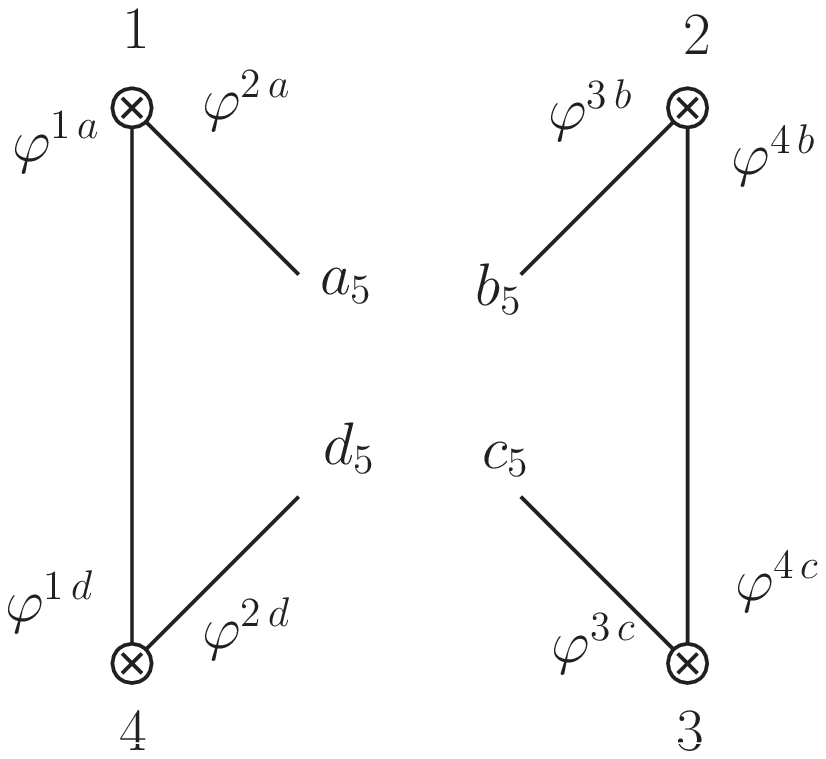} }
\qquad\qquad \subfloat[]{\label{fig:disc-quart-split2}
\includegraphics[width=0.2\textwidth]{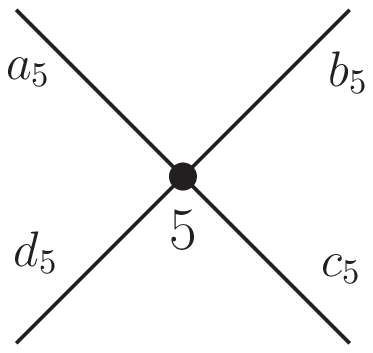} }
\end{center}
\caption{Factorisation of diagram involving a quartic vertex.}
\label{fig:disc-quartic-split}
\end{figure}

We factorise the diagram as in figure
\ref{fig:disc-quartic-split}. The different Wick contractions
correspond to inequivalent ways of gluing together parts (a) and (b)
in the figure. 

The following contribution comes from figure
\ref{fig:disc-quart-split1} and is common to all diagrams in this set
\ba
\label{common-1}
&&\hsp{-1} E_4[a_5,b_5,c_5,d_5] = \fr{16}\left(\fr{8}\right)^{\!4}
\sigma^{1m_1n_1}\sigma^{2p_1q_1}\sigma^{3p_2q_2}
\sigma^{4m_2n_2} \sigma^{3p_3q_3}\sigma^{4m_3n_3}
\sigma^{1m_4n_4}\sigma^{2p_4q_4}\;  k^6 \delta^{ad}\delta^{bc}  \nn \\
&& \hsp{-0.5}\times \!\left(\frac{i}{\sqrt{2}}\right)^{\!8}
\left( \bar d_{m_1} \bar d_{n_1} \la d^4\ra 
\frac{\delta^8_{14}}{x^2_{14}} \overleftarrow{\bar d_{n_4} 
\bar d_{m_4}} \right) \!\!\left( \bar d_{m_2} \bar d_{n_2} \la d^4\ra 
\frac{\delta^8_{23}}{x^2_{23}} \overleftarrow{\bar d_{n_3} 
\bar d_{m_3}} \right) \!f^{ea_5b_5}\,f^{ec_5d_5}
\left(-\frac{g^2}{64}\right).
\ea
This common portion simplifies to
\be
\label{common-2}
\hsp{-0.25}E_4[a_5,b_5,c_5,d_5] = T(\sigma)\!\!
\left(-\frac{g^2}{2^{26}}\right)\!k^6\,\d^{ad}\d^{bc} 
f^{ea_5b_5}\,f^{ec_5d_5} \frac{(4!)^6}{8^4} \veps_{m_1n_1n_4m_4} 
\veps_{m_2n_2n_3m_3} \fr{x^2_{14}}\fr{x^2_{23}},
\ee
where $T(\sigma)$ denotes the product of the eight $\sigma$
coefficients in (\ref{common-1}).

We now need to consider all possible ways of gluing of this factor
with the piece resulting from figure \ref{fig:disc-quart-split2}. We
use the following notation, \\
\hsp{1} \raisebox{-20pt}{
\includegraphics[width=0.15\textwidth]{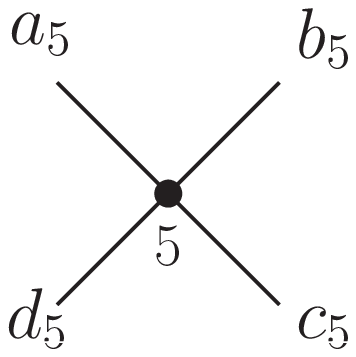}} 
$\equiv$ $V_4[a_5,b_5,c_5,d_5]$\,, \\
where the order of the arguments in $V_4$ \rule{0pt}{22pt}corresponds
to the clockwise labelling in the vertex starting from the top left
leg.

The different Wick contractions are analysed in appendix
\ref{app:rules-quartic}. Combining all the non-zero contributions we
find that figure \ref{fig:disc-quartic} evaluates to
\be
-g^2 f^{abc} f^{abc} \frac{1}{8(2\pi)^{12}} \fr{x^2_{14}x^2_{23}}
\int \dr^4 x_5 \fr{x^2_{51}x^2_{52}x^2_{53}x^2_{54}}\,.
\ee
Using $f^{abc}f^{abc}=N(N^2-1)$ and including all multiplicity factors 
the complete one-loop contribution to (\ref{G4H-def}) is therefore

\vsp{0.3}

\be
\left[G_4^{(H)}(x_1,\ldots,x_4)\right]_{\rm 1-loop} = 
-g^2N(N^2-1) \frac{1}{2(2\pi)^{12}} \fr{x^2_{14}x^2_{23}}
\int \dr^4 x_5 \fr{x^2_{51}x^2_{52}x^2_{53}x^2_{54}}\,.
\label{G4H-1loop}
\ee

\vsp{0.6}

\ndt
The box integral in (\ref{G4H-1loop}) is well known \cite{box} and can
be expressed in terms of the cross ratios (\ref{cross-ratios}).  Using
the form of the box integral in \cite{bkrs-1}, the one-loop
contribution to $G_4^{(H)}(x_1,\ldots,x_4)$ takes the form
\be
\left[G_4^{(H)}(x_1,\ldots,x_4)\right]_{\rm 1-loop} = 
-g^2N(N^2-1)\pi^2 \frac{1}{2(2\pi)^{12}} 
\fr{x^2_{14}x^2_{23}x^2_{13}x^2_{24}}\,
F_4^{(H)}(r,s) \, ,
\label{G4H-1loop-rs}
\ee
where $F_4^{(H)}(r,s)$ can be expressed as a combination 
of logarithms and dilogarithms as 
\ba
F_4^{(H)}(r,s) & = & 
{1 \over \sqrt{p}}
\left \{ \log (r)\log (s)  - 
\left [\log \left({r+s-1 -\sqrt {p} \over 2}\right)\right ]^{2} + 
\right. \nonumber \\ 
&& \left. -2 \,{\rm Li}_2 \left({2 \over 1+r-s+\sqrt {p}}\right ) -
2 \,{\rm Li}_2 \left({2 \over 1-r+s+\sqrt {p}}\right )\right \} 
\; ,  \label{Brsf}
\ea
where ${\rm Li}_2 (z) = \sum_{n=1}^{\infty} {z^n\over n^2}$ 
and
\be 
p = 1 + r^{2} + s^{2} - 2r - 2s - 2rs \, .  
\label{pdef}
\ee

\section{Open problems and future directions}

In this paper we have initiated a program aimed at systematically
studying correlation functions of gauge-invariant operators in
$\calN=4$ SYM using the light-cone superspace formulation. Our main
goals are on the one hand to develop efficient techniques for the
computation of perturbative corrections to correlation functions and
on the other to shed light on subtleties which can potentially arise from
the use of the light-cone gauge in the calculation of off-shell
quantities.  

As a computational tool light-cone superspace is particularly
promising for a number of reasons. This formulation of the $\calN=4$
SYM theory uses only one type of superfield, which carries no
space-time or SU(4)$_R$ indices. Therefore the general structure of
super Feynman diagrams and the combinatorial analysis involved in
their study are simpler than in other formulations. Moreover, while
the one-loop calculation we presented did not show noticeable
simplifications compared to similar covariant calculations, we expect
that the manifest  $\calN=4$ supersymmetry will lead to a significant
computational advantage, in terms of the number of diagrams to
evaluate, at higher orders in the perturbative expansion. 

Particularly interesting is the possibility of taking advantage of the
full $\calN=4$ supersymmetry to describe in compact form entire
multiplets of operators and their correlation functions. This should
be possible for the multiplet of the energy-momentum tensor, which is
expected to take the form of a linear combination of quadratic, cubic
and quartic terms in the $\calN=4$ superfield, $\Phi$. In order to
determine the exact combination as well as to generalise such a
construction to different multiplets, it will be important to better
understand the role played by the non-linearly realised dynamical
supersymmetries. 

In the case of the simple four-point function
$G_4^{(H)}(x_1,\ldots,x_4)$ we reproduced the known result to one-loop
order. The light-cone gauge thus yields a manifestly Lorentz covariant
result. This is thanks to non-trivial cancellations of derivatives and
$1/\del_-$ factors. It will be important to understand these
cancellations in a systematic way for more complicated correlation
functions and/or at higher orders in perturbation theory. 

A distinct, but related, issue concerns the general consistency of the
light-cone gauge formalism, in its superspace realisation, when
applied to the study of off-shell observables. In the case of a
(super) conformal gauge theory such as $\calN=4$ SYM the potential
subtleties are associated with spurious infra-red divergences induced
by the presence of the $1/\del_-$ operators. In the case of the simple
four-point function that we studied in this paper, various
cancellations ensured the absence of any such singularities from the
final result. At this stage we do not yet have a clear understanding
of how (or even if) similar cancellations take place in general
perturbative calculations. The question of whether or not spurious
infra-red divergences arise in generic gauge-invariant correlation
function is therefore still open and this is an aspect that deserves
further consideration. 

Another important point that remains to be addressed is the
identification of the most convenient regularisation method to deal
with divergent integrals. Previous applications of light-cone
superspace, including the all order proof the ultra-violet finiteness
of $\calN=4$ SYM, did not require the use of an explicit
regularisation. However, divergences do arise in the calculation of
correlation functions of non-protected operators such as the Konishi
operator (\ref{d2-konishi})-(\ref{superK}). Hence a suitable
regularisation scheme will be needed for such calculations. As an
added benefit this will also make it possible to explicitly prove the
non-renormalisation of two- and three-point functions of protected
operators.

We consider the results presented in this paper to be encouraging and
we hope to address the open questions outlined above in future
publications.

\vskip 0.7cm
\ndt 
{\bf {Acknowledgments}}\\[0.3cm]
We thank Y.S. Akshay, L. Brink and H. Shimada for valuable discussions. This work is supported by the Max Planck Society, Germany, through the Max
Planck Partner Group in Quantum Field Theory. S.A. acknowledges
support by the Department of Science and Technology, Government of
India, through a Ramanujan Fellowship. S.P. is supported by a summer
research fellowship from the Indian Academy of Sciences, Bangalore,
and an INSPIRE grant from the Department of Science and Technology,
Government of India.

\vskip 1cm
\appendix

\section{Conventions and notation}
\label{app:conventions}

We work with space-time signature $(-,+,+,+)$ and define the
light-cone coordinates and their derivatives as
\bea
&& \hsp{-1}x^{\pm} = \frac{1}{\sqrt 2}({x^0}\,{\pm}\,{x^3})\, , \quad
x = \frac{1}{\sqrt 2}({x^1}+i\,{x^2}) \, , \quad 
{\bar x} = \frac{1}{\sqrt 2}({x^1}-i\,{x^2})\, , 
\label{lc-ccord} \\
&& \hsp{-1} {\partial_{\pm}}=\frac{1}{\sqrt 2}({\partial_0}\,{\pm}\,
{\partial_3})\,, \quad {\bar\partial} =\frac{1}{\sqrt 2}
({\partial_1}-i\,{\partial_2})\, , \quad
{\partial} =\frac{1}{\sqrt 2}({\partial_1}+i\,{\partial_2})\, .
\label{lc-deriv}
\eea
The gauge field components in the light-cone decomposition are
\be
A_\pm = \fr{\sqrt{2}}(A_0\pm A_3) \, , \quad A = \fr{\sqrt{2}} 
( A_1+ i A_2)\, , \quad \bar A = \fr{\sqrt{2}} ( A_1 - i A_2) \, .
\ee
The light-cone gauge fixing involves setting $A_-=0$ and integrating
out $A_+$, leaving only the two transverse physical components, $A$
and $\bar A$.  The four Weyl fermions in the $\calN=4$ multiplet,
$\psi^m_\a$, and their conjugates, $\bar\psi_{m\,\adot}$, are
decomposed according to the projection
\be
\psi^m_\a ~ \to ~ \psi_{(\pm)}^m = \calP_\pm \psi^m_\a \, , \qquad
\bar\psi_{m\,\adot} ~ \to ~ \bar\psi^{(\pm)}_m = 
\calP_\pm \bar\psi_{m\,\adot} \, ,
\ee
where $\calP_\pm = -\fr{\sqrt{2}} (\s^0\pm\s^3)$. The $\psi^m_{(+)}$
and $\bar\psi_m^{(+)}$ components can be integrated out. The
light-cone description uses the remaining one-component fermionic
fields $\lambda^m\equiv\psi^m_{(-)}$, and their conjugates,
$\bar\lambda_m\equiv\bar\psi^{(-)}_m$. 

The Grassmann integrals in light-cone superspace are normalised so that 
\be
\int \dr\th_m\,\th^n = \delta_m^n \, , \qquad 
\int \dr\bth^m\,\bth_n = \d^m_n \, .
\label{grassint-1}
\ee
We define
\be
\dr^4\th = \fr{(4!)^2} \,\veps^{mnpq}\dr\th_m\dr\th_n\dr\th_p\dr\th_q \,,
\qquad \dr^4\bth = \fr{(4!)^2} \,\veps_{mnpq}\dr\bth^m\dr\bth^n
\dr\bth^p\dr\bth^q \, .
\ee
This, together with (\ref{grassint-1}), ensures that 
\be
\int \dr^4\th\,\delta^{(4)}(\th) = \int \dr^4\bth\,\delta^{(4)}(\bth) = 1 \, ,
\ee
where the $\delta$-functions are defined as
\be
\delta^{(4)}(\th) = \la\th^4\ra \equiv \veps_{mnpq}\th^m\th^n\th^p\th^q
\, , \qquad \delta^{(4)}(\bth) = \la\bth^4\ra \equiv \veps^{mnpq}
\bth_m\bth_n\bth_p\bth_q \, .
\label{fermidelta}
\ee

\vsp{0.3}
\ndt
The scalar fields in the $\calN=4$ multiplet can be represented
either as SU(4)$_R$ bi-spinors, $\v^{mn}$, satisfying the reality condition
(\ref{reality}) or as vectors, $\v^i$, $i=1,\ldots,6$. The
two representations are related by 
\be
\v^i=\fr{\sqrt 8}\Sigma^i_{mn}\v^{mn}=\fr{2\sqrt 8}\,\veps^{mnpq}
\Sigma^i_{mn}\,\bar\v_{pq}=\fr{\sqrt 8}\,\sigma^{i\,pq}\,\bar\v_{pq}\ .
\label{so6su4}
\ee
where $\S^i_{mn}$ ($\bar\S^{mn}_i$) are Clebsch-Gordan coefficients
relating the product of two $\bf{4}$'s ($\bf{\bar 4}$'s) to the
$\bf{6}$ of SU(4). They are defined as follows
\ba
&& \hsp{-1} \S^i_{mn} = (\S^I_{mn},\S^{I+3}_{mn}) = 
(\eta^I_{mn},i\bar\eta^I_{mn}) \, ,  \nn \\
&& \hsp{-1} \bar\S^{mn}_i = (\bar\S^{mn}_I,\bar\S^{mn}_{I+3}) = 
(\eta^I_{mn},-i\bar\eta^I_{mn}) \, , \qquad I=1,2,3 \, ,
\label{446-CGcoeffs}
\ea
where $(\eta^I_{mn},i\bar\eta^I_{mn})$ are 't Hooft symbols,
\ba
&& \hsp{-1} \eta^I_{mn} = \bar\eta^I_{mn} = \veps_{Imn} \, , 
\quad m,n=1,2,3 \nn \\
&& \hsp{-1} \eta^I_{m4} = \bar\eta^I_{4m} = \d^I_m \, , 
\quad m=1,2,3 \nn \\
&& \hsp{-1} \eta^I_{mn} = -\eta^I_{nm} \, , \quad \bar\eta^I_{mn} = 
-\bar\eta^I_{nm} \, .
\ea
Splitting up the $i$ index in terms of $I=1,2,3$, the coefficients 
(\ref{446-CGcoeffs}) can be written as
\bea
\Sigma^I_{mn}&=&\veps^I_{\;mn4}+(\delta^I_m\delta^4_n-\delta^I_n
\delta^4_m)\, , \nn \\
\Sigma^{I+3}_{mn}&=&i\veps^I_{\;mn4}-i(\delta^I_m\delta^4_n-
\delta^I_n\delta^4_m)\, .
\eea
From this we obtain the $\s^{i\,mn}$ coefficients
\bea
\sigma^{Ipq}&=&\veps^{Ipq4}+(\delta^{Ip}\delta^q_4-
\delta^p_4\delta^{Iq})\, , \nn \\
\sigma^{(I+3)pq}&=&-i\veps^{Ipq4}+i(\delta^{Ip}\delta^q_4-
\delta^p_4\delta^{Iq})\ .
\label{def:sigma}
\eea

\section{Derivation of super-propagator}
\label{app:prop}

In this appendix we discuss in detail the derivation of the propagator
(\ref{super-prop}) for the $\calN=4$ superfield. We start with a path
integral derivation which will allow us to check the consistency of
various conventions for Grassmann integrals and functional
derivatives. 

\subsection{Path integral derivation}
\label{app:prop1}

The superfield propagator can be obtained inverting the kinetic
operator in (\ref{N4action-2}). We can obtain it constructing the
generating functional for Green functions of the $\calN=4$ superfield
in the free theory limit, $Z_0[J]$. 

Functional differentiation of $Z[J]$ with respect to the sources,
$J(x,\th,\bth)$, gives rise to Green functions of the $\calN=4$
superfields. Because of the chirality of both $\Phi$ and $J$ we need
to be careful in defining the rules for functional differentiation in
superspace. In defining the functional derivative with respect to a
chiral superfield we require the condition that the variation of a
chiral superfield be chiral. To satisfy this condition we consider a
chiral superfield, $\Psi(x,\th,\bth)$, written in terms of the chiral
variable (\ref{chirvar}) and we impose
\be
\frac{\delta\Psi(y',\th')}{\delta\Psi(y,\th)} = \delta^{(4)}(y-y')
\delta^{(4)}(\th-\th') \, .
\label{functder-1}
\ee
To obtain the form of the derivative
$\d\Psi(x',\th',\bth')/\d\Psi(x,\th,\bth)$ in terms of the standard
superspace coordinates we consider
\be
\frac{\delta}{\delta\Psi(x,\th,\bth)} \int\dr^4x'\dr^4\th'\dr^4\bth'\,
\Psi(x',\th',\bth') \, F(x',\th',\bth') \, ,
\label{functder-2}
\ee
where $F(x,\th,\bth)$ is a generic (non-chiral) superfield. Using
(\ref{functder-1}) we can evaluate (\ref{functder-2}) as follows
\ba
&& \hsp{-1} \frac{\delta}{\delta\Psi(x,\th,\bth)} \int\dr^4x'
\dr^4\th'\dr^4\bth'\,\Psi(x',\th',\bth') \, F(x',\th',\bth') \nn \\
&& \hsp{-1} = \int \dr^4y'\dr^4\th'\dr^4\bth' \,
\frac{\d\Psi(y',\th')}{\d\Psi(y,\th)}
\,F({x'}^+,{y'}^-+\frac{i}{\sqrt{2}}\th'\bth',x',\bar x',\th',\bth') \nn \\
&& \hsp{-1} = \int \dr^4\bth' \, F(x^+,y^-+
\frac{i}{\sqrt{2}}\th\bth',x,\bar x,\th,\bth') 
\nn \\
&& \hsp{-1} = \fr{(4!)^2} \la d^4 \ra F(x,\th,\bth) \, ,
\label{functder-3}
\ea
where in the last step we used 
\be
\int \dr\bth^k \, F(x^+,y^-+\frac{i}{\sqrt{2}}\th\bth,x,\bar x,\th,\bth) = 
d^k F(x^+,x^-,x,\bar x,\th,\bth) \, , \qquad k=1,\ldots,4 \, ,
\ee
which can be verified expanding left and right hand sides in
components. From (\ref{functder-3}) we deduce the rule for functional
differentiation with respect to a chiral superfield,
\be
\frac{\d\Psi^a(x',\th',\bth')}{\d\Psi^b(x,\th,\bth)} = \fr{(4!)^2} \d^a_b
\la d^4 \ra \d^{(4)}(x-x')\d^{(4)}(\th-\th')\d^{(4)}(\bth-\bth') \, ,
\label{chi-functder}
\ee
which applies in particular to the $\calN=4$ superfield, $\Phi$. For
its conjugate, $\bar\Phi$, using (\ref{Phibar}), we get
\be
\frac{\d\bar\Phi^a(x',\th',\bth')}{\d\Phi^b(x,\th,\bth)} = 
\fr{2(4!)^3} \d^a_b\frac{\la\bar d^4 \ra \la d^4 \ra}{\del_-^2} 
\d^{(4)}(x-x')\d^{(4)}(\th-\th')\d^{(4)}(\bth-\bth') \, .
\ee
We can now define the generating functional, $Z[J]$, as follows
\be
Z[J] = \frac{\displaystyle\int[\dr\Phi]\,\er^{-\calS[\Phi]+
\int\dr^{12}z\,\Phi^a(z) \frac{\la \bar d^4 \ra}{4\del_-^4} J_a(z)}}
{\displaystyle\int[\dr\Phi]\,\er^{-\calS[\Phi]}} \, , 
\label{gen-functnl-1a}
\ee
where, as usual, $\dr^{12}z=\dr^4x\,\dr^4\th\,\dr^4\bth$. 

Notice, in particular, the coupling to the sources, $J(z)$, in
(\ref{gen-functnl-1a}). This is chosen so as to produce the correct
coupling to external sources in the equations of motion. This can be
seen considering the free theory in the presence of external sources,
\be
\int \dr^{12}z\, \half\Phi^a(z)\,\calK_a^{\;b}\,\Phi_b(z) + \int \dr^{12}z \, 
\Phi^a(z) \frac{\la \bar d^4 \ra}{4\del_-^4} J_a(z) \, , 
\label{free-th+source}
\ee
where the kinetic operator is 
\be
\calK_a^{\;b} = -3\, \d_a^{\;b}\,\frac{\la \bar d^4 \ra \Box}{\del_-^4} \, .
\label{kinop}
\ee
Varying (\ref{free-th+source}) with respect to the superfield $\Phi$
gives rise to the correct equations of motion in the presence of an
external source,
\be
\fr{(4!)^2} \la d^4 \ra \calK_a^{\;b}\,\Phi_b(x,\th,\bth) = J_a(x,\th,\bth) \, .
\ee
The right hand side is straightforward to obtain using the definition
(\ref{chi-functder}), 
\ba
&& \hsp{-1} \frac{\d}{\d\Phi^a(z)} \int\dr^{12}z'\,\Phi^b(z')
\frac{\la \bar d^4 \ra}{4\del_-^4} J_b(z') = 
\fr{(4!)^2} \int \dr^{12}z' \, \la d^4 \ra \d^{(12)}(z-z') 
\frac{\la \bar d^4 \ra}{4\del_-^4} J_a(z') \nn \\
&& \hsp{-1} = \fr{(4!)^2} \int \dr^{12}z' \, \d^{(12)}(z-z') 
\frac{\la d^4 \ra \la \bar d^4 \ra}{4\del_-^4} J_a(z')
= \int \dr^{12}z' \, \d^{(12)}(z-z') J_a(z') = J_a(z) \, ,
\ea
where we used the fact that $\la d^4 \ra\la \bar d^4 \ra = 4
(4!)^2\,\del_-^4$ when acting on a chiral superfield such as $J(z)$. 

In the free theory limit the exponent in the generating functional
(\ref{gen-functnl-1a}) reduces to
\be
-\half \left(\!\Phi^a,\calK_a^{\;b}\,\Phi_b\!\right) + 
\left(\!\Phi^a,\frac{\la \bar d^4 \ra}{4\del_-^4} J_a\!\right)
= -\half \int \dr^{12}z \,\Phi^a(z) \, \calK_a^{\;b}\,\Phi_b(z)  +
\int\dr^{12}z\,\Phi^a(z) \frac{\la \bar d^4 \ra}{4\del_-^4} J_a(z) \, .
\ee
The functional integral (\ref{gen-functnl-1a}) becomes Gaussian and
thus straightforward to compute. The result is
\be
Z_0[J] = \er^{\half\left(\wtilde J^a,[\calK^{-1}]_a^{\;b}\,\wtilde J_b\right)}
\label{free-gen-functnl-2a}
\ee
where 
\be
\wtilde J^a(z) = \frac{\la \bar d^4 \ra}{4\del_-^4} J^a(z) 
\ee
and $\calK^{-1}$ is the inverse of the kinetic operator
(\ref{kinop}). In (\ref{free-gen-functnl-2a}) a factor of
$\det(\calK)^{-1/2}$ has been cancelled between numerator and
denominator. The free generating functional (\ref{free-gen-functnl-2a})
allows to construct the perturbative expansion of the full functional
$Z[J]$ in (\ref{gen-functnl-1}).  

Introducing the kernel, $\D(z,z')$, of the operator $\calK^{-1}$, we
can rewrite (\ref{free-gen-functnl-2a}) as
\be
Z_0[J] = \er^{\half\int\dr^{12}z\,\dr^{12}z' \: \wtilde J^a(z) 
[\D(z,z')]_a^{\;b}\wtilde J_b(z')} \, .
\label{free-gen-functnl-3}
\ee
$\D(z,z')$ is of course the super-propagator we are interested in. 
Let us denote by $K(z,z')$ the kernel of the kinetic operator (\ref{kinop}),
\be
K(z,z') = -3\,\d^{(12)}(z-z')\,\frac{\la \bar d^4 \ra \Box}{\del_-^4} \, ,
\label{kinkernel}
\ee
where $\d^{(12)}(z-z')=\d^{(4)}(x-x')\d^{(4)}(\th-\th')\d^{(4)}(\bth-\bth')$. 
Then $\D(z,z')$ is defined by the condition
\be
\int \dr^{12}z'' \, \D(z,z'')\,K(z'',z') = \d^{(12)}(z-z') \, ,
\ee
or, introducing a chiral test superfield, $\Psi(z)$, 
\be
\int \dr^{12}z'' \int \dr^{12}z'\, \D(z,z'')\,K(z'',z')\,\Psi(z') 
= \Psi(z) \, .
\ee
Using the explicit form (\ref{kinkernel}) of $K(z,z')$ we have 
\ba
\Psi(z) &\!\!=\!\!& \int \dr^{12}z' \int \dr^{12}z'' \, \D(z,z'')
\d^{(12)}(z''-z')\left(-3\frac{\la \bar d^4 \ra\Box}{\del_-^4}
\Psi\right)\!(z') \nn \\
&\!\!=\!\!& \int \dr^{12}z' \, \D(z,z') \left(-3\frac{\la \bar d^4 \ra\Box}
{\del_-^4}\Psi\right)\!(z') \, .
\label{inv-kernel-1}
\ea
The solution for $\D(z,z')$ is of the form
\be
\D(z,z') = k\,\frac{\la d^4 \ra}{(x-x')^2} \d^{(4)}(\th-\th')
\d^{(4)}(\bth-\bth') \, ,
\ee
with $k$ a constant to be fixed. Substituting into the right hand side of 
(\ref{inv-kernel-1}) we get
\ba
&& \hsp{-1} \int \dr^{12}z' \, k\,\frac{\la d^4 \ra}{(x-x')^2} 
\d^{(4)}(\th-\th')\d^{(4)}(\bth-\bth') \left(-3\frac{\la \bar d^4 \ra\Box}
{\del_-^4}\Psi\right)\!(z') \nn \\
&& \hsp{-1} = -3k\int \dr^{12}z' \, \Box\fr{(x-x')^2} \d^{(4)}(\th-\th')
\d^{(4)}(\bth-\bth') \left(\frac{\la d^4 \ra\la \bar d^4 \ra}
{\del_-^4}\Psi\right)\!(z') \nn \\
&& \hsp{-1} = -3 k (2\pi)^2 4 (4!)^2 \int \dr^{12}z'\,
\d^{12}(z-z') \Psi(z') = -\frac{3k(4!)^3(2\pi)^2}{2} \, \Psi(z) \, ,
\label{inv-kernel-2}
\ea
where we used integration by parts and the relations 
\be
\Box\fr{(x-x')^2} = (2\pi)^2\d^{(4)}(x-x') 
\ee
and
\be
\la d^4 \ra\la \bar d^4 \ra \Psi(z) = 4(4!)^2\del_-^4 \Psi(z) \, .
\ee 
The latter is valid for a chiral superfield $\Psi(z)$. From
(\ref{inv-kernel-2}) we read off the value of the constant $k$,
\be
k = -\frac{2}{(4!)^3(2\pi)^2} \, .
\ee
So the superfield propagator is
\be
\Delta^a_b(z-z') = 
-\frac{2}{(4!)^3}\frac{\d^a_b}{(2\pi)^2} \fr{(x-x')^2} \la d^4 \ra
\d^{(4)}(\th-\th')\d^{(4)}(\bth-\bth') \, .
\label{super2pt-fin}
\ee

\subsection{Relation to component field propagators}
\label{app:prop2}

In order to verify that the superfield propagator constructed in the
previous subsection contain the correct propagators for the individual
fields in the $\calN=4$ multiplet we now re-derive the
$\D(z-z^\prime)$ starting from the component expansion of $\Phi(z)$. 

In the following it will be convenient to rewrite the $\calN=4$ superfield 
(\ref{N4supfield-1}) as
\ba
\Phi\,(x,\theta,\bar\theta)&\!\!=\!\!&\er^{
-\frac{i}{\sqrt{2}}\th^m\bth_m\del_-}\left[-\frac{1}{\del_-}A(x)
-\frac{i}{\del_-}\theta^m{\bar \lambda}_m(x)
+\frac{i}{\sqrt 2}\,\theta^m\theta^n{\bar \v}_{mn}(x)\right.\nn \\
&&\left.+\frac{\sqrt 2}{6}\,\veps_{mnpq}\theta^m\theta^n\theta^p
\lambda^q(x) -\frac{1}{12}\,\veps_{mnpq}\theta^m\theta^n\theta^p
\theta^q\,\del_-{\bar A}(x) \right] \, .
\label{N4supfield-2}
\ea
The kinetic terms in the $\calN=4$ light-cone component action are 
\be
S_0 = \int \dr^4 x \, \left[ \bar A(x)\Box A(x) + \half\v_i(x)\Box\v^i(x)
-\frac{i}{\sqrt{2}}\bar\lambda_m(x)\frac{\Box}{\del_-}\lambda^m(x)
\right] \, ,
\label{freecompaction}
\ee
where the relation between the six real scalar fields $\v^i$,
$i=1,\ldots,6$ and the $\v^{mn}$'s, $m,n=1,\ldots,4$ in
(\ref{N4supfield-2}) involves Clebsch-Gordan coefficients and it is
given explicitly in (\ref{so6su4}).

From (\ref{freecompaction}) we get the free propagators for the
component fields,
\ba
&& \hsp{-1.5} \left(\D^{\!\scriptscriptstyle(A)}\right)^a_b(x-y) 
= \la \bar A^a(x) \,A_b(y) \ra = \fr{(2\pi)^2}\,\frac{\d^a_b}{(x-y)^2} 
\label{gauge-prop} \\
&& \hsp{-1.5} \left(\D^{\!\scriptscriptstyle(\v)}\right)^{a\,ij}_{b}(x-y) 
= \la \v^{a\,i}(x) \,\v^j_b(y) \ra = 
\fr{(2\pi)^2}\,\frac{\d^{ij}\d^a_b}{(x-y)^2} 
\label{scalar-prop} \\ 
&& \hsp{1.36} \Rightarrow ~
\left(\D^{\!\scriptscriptstyle(\v)}\right)^{a\,pq}_{b\,mn}(x-y) 
= \la \bar\v_{a\,mn}(x)\,\v_b^{pq}(y) \ra = \fr{(2\pi)^2}\,
\frac{(\d_m^q\d_n^p-\d_m^p\d_n^q)\d^a_b}{(x-y)^2} \nn \\
&& \hsp{-1.5} \left(\D^{\!\scriptscriptstyle(\lambda)}
\right)^{a\,n}_{b\,m}(x-y) 
= \la \bar\lambda^a_m(x) \,\lambda_b^n(y) \ra = 
\frac{i\sqrt{2}}{(2\pi)^2}\, \del_-\frac{\d_m^n\d^a_b}{(x-y)^2} = 
\frac{i\sqrt{2}}{(2\pi)^2}\,
\frac{\d_m^n\d^a_b(x^+-y^+)}{(x-y)^4}  \, .
\label{fermion-prop}
\ea
We can now consider the superfield two-point function,
\be
\D^a_b(x,\th,\bth;x^\prime,\th^\prime,\bth^\prime) = 
\la \Phi^a(x,\th,\bth)\,\Phi_b(x^\prime,\th^\prime,\bth^\prime) \ra \, .
\label{super2pt-0}
\ee
Using (\ref{N4supfield-2}), we expand this two-point function as
\ba
&& \hsp{-1.45} \la \Phi^a(x,\th,\bth)\,
\Phi_b(x^\prime,\th^\prime,\bth^\prime) \ra
= \er^{-\frac{i}{\sqrt{2}}(\th^m\bth_m\del_- + 
{\th^\prime}^m{\bth^\prime}_m\del_-^\prime)} 
\la \left[-\frac{1}{\del_-}A^a(x)-\frac{i}{\del_-}\theta^m
{\bar \lambda^a}_m(x) \right. \nn \\
&& \left.+\frac{i}{\sqrt 2}\,\theta^m\theta^n{\bar \v^a}_{mn}(x)
+\frac{\sqrt 2}{6}\,\veps_{mnpq}\theta^m\theta^n\theta^p
\lambda^{a\,q}(x) -\frac{1}{12}\,\veps_{mnpq}\theta^m\theta^n\theta^p
\theta^q\,\del_-{\bar A}^a(x) \right] \nn \\
&& \left[-\frac{1}{\del_-^\prime}A_b(x^\prime) -\frac{i}{\del_-^\prime}
{\theta^\prime}^r{\bar\lambda}_{b\,r}(x^\prime) 
+ \frac{i}{2\sqrt 2}\,\veps_{rsuv}{\theta^\prime}^r{\theta^\prime}^s
\v^{uv}_b(x^\prime)+\frac{\sqrt 2}{6}\,\veps_{rsuv}{\theta^\prime}^r
{\theta^\prime}^s{\theta^\prime}^u\lambda^v_b(x^\prime) \right. \nn \\
&& \left. -\frac{1}{12}\,\veps_{rsuv}
{\theta^\prime}^r{\theta^\prime}^s{\theta^\prime}^u
{\theta^\prime}^v \del_-^\prime{\bar A}_b(x^\prime) \right] \ra \, ,
\label{super2pt-1}
\ea
where $\del_-^\prime=\del/\del {x^\prime}^-$ and we used the reality
condition 
\be
\bar\v_{mn}(x) = \half\,\veps_{mnpq}\,\v^{pq}(x)
\label{realscalar}
\ee
for the scalar field in the second superfield. 

In the superspace two-point function (\ref{super2pt-1}) the only
non-zero contractions correspond to the component two-point functions
(\ref{gauge-prop})-(\ref{fermion-prop}). Therefore we get
\ba
&& \hsp{-1} \la \Phi^a(x,\th,\bth)\,
\Phi_b(x^\prime,\th^\prime,\bth^\prime) \ra
= \er^{-\frac{i}{\sqrt{2}}(\th^m\bth_m \del_- + 
{\th^\prime}^m{\bth^\prime}_m\del_-^\prime)} \left[ \fr{12}
\veps_{mnpq} \th^m\th^n\th^p\th^q \la\del_-\bar A^a(x)\,
\fr{\del_-^\prime} A_b(x^\prime) \ra 
\right. \nn \\
&& +\fr{12} \veps_{mnpq} {\theta^\prime}^m{\theta^\prime}^n
{\theta^\prime}^p {\theta^\prime}^q \la \fr{\del_-}A^a(x) \,
\del_-^\prime\bar A_b(x^\prime) \ra -\fr{4} \veps_{mnpq} \th^r\th^s
{\th^\prime}^m{\th^\prime}^n\la\bar\v^a_{rs}(x)\,\v^{pq}_b(x^\prime)\ra 
\label{super2pt-2} \\
&& \left.-i\frac{\sqrt{2}}{6}\veps_{mnpq}\th^r{\th^\prime}^m{\th^\prime}^n
{\th^\prime}^p\la\fr{\del_-}{\bar\lambda}_r^a(x)\,\lambda^q_b(x^\prime)\ra
-i\frac{\sqrt{2}}{6}\veps_{mnpq}\th^m\th^n\th^p{\th^\prime}^r
\la\lambda^{q\,a}(x)\,\fr{\del_-}{\bar\lambda}_{r\,b}(x^\prime)\ra
\right] \nn \, ,
\ea
Using (\ref{gauge-prop})-(\ref{fermion-prop}) and integration by parts
to get rid of the extra $\del_-$'s, we find
\ba
&& \hsp{-0.3}
\la \Phi^a(x,\th,\bth)\,\Phi_b(x^\prime,\th^\prime,\bth^\prime) \ra
= \delta^a_b\, \er^{-\frac{i}{\sqrt{2}}(\th^m\bth_m - 
{\th^\prime}^m{\bth^\prime}_m)\del_-} \veps_{mnpq}\left[-\fr{12}
\th^m\th^n\th^p\th^q - \fr{12}{\th^\prime}^m{\th^\prime}^n{\th^\prime}^p
{\th^\prime}^q \right. \nn \\
&& \hsp{3.8}\left.- \half \th^m\th^n{\th^\prime}^p{\th^\prime}^q 
+\fr{3} \th^m{\th^\prime}^n{\th^\prime}^p{\th^\prime}^q
+\fr{3} \th^m\th^n\th^p{\th^\prime}^q \right]
\fr{(2\pi)^2}\,\fr{(x-x^\prime)^2} \nn \\
&& \hsp{3.87} = -\fr{12(2\pi)^2} \,\delta^a_b\,  \er^{-\frac{i}{\sqrt{2}}
(\th^m\bth_m - {\th^\prime}^m{\bth^\prime}_m)\del_-} 
\frac{\delta^{(4)}(\th-\th^\prime)}{(x-x^\prime)^2} \, .
\label{super2pt-3}
\ea
where we used the definition (\ref{fermidelta}) of the fermionic
$\delta$-function. The super-propagator can be put in a more
convenient form using the following identity
\be
\la d^4 \ra \delta^{(4)}(\bth-\bth^\prime) = (4!)^2 \, \er^{
-\frac{i}{\sqrt{2}}(\th^m\bth_m - \th^m{\bth^\prime}_m)\del_-} \, ,
\label{d4bdelta4}
\ee
which can be proven expanding the left hand side as
\ba
\la d^4 \ra \delta^{(4)}(\bth-\bth^\prime) &\!\!=\!\!& \veps_{mnpq}
\veps^{rsuv}d^md^nd^pd^q(\bth_r-\bth^\prime_r)(\bth_s-\bth^\prime_s)
(\bth_u-\bth^\prime_u)(\bth_v-\bth^\prime_v) \nn \\
&\!\!=\!\!& (4!)^2 \,d^1d^2d^3d^4 (\bth_1-\bth^\prime_1)
(\bth_2-\bth^\prime_2)(\bth_3-\bth^\prime_3)(\bth_4-\bth^\prime_4)
\rule{0pt}{14pt} 
\ea
and using (no sum over the repeated index $k$)
\be
d^k(\bth_k-\bth^\prime_k) = -1 + \frac{i}{\sqrt{2}} (\th^k\bth_k
- \th^k\bth^\prime_k)\del_- = -\er^{-\frac{i}{\sqrt{2}}(\th^k\bth_k-
\th^k\bth^\prime_k)\del_-} \, \qquad k=1,\ldots,4  \, .
\ee
The identity (\ref{d4bdelta4}) can be rewritten as
\be
1 = \fr{(4!)^2} \, \er^{+\frac{i}{\sqrt{2}}(\th^m\bth_m-
\th^m\bth^\prime_m)\del_-} \la d^4 \ra \delta^{(4)}(\bth-\bth^\prime) \, .
\label{1-identity}
\ee
Inserting (\ref{1-identity}) into the expression for the
super-propagator we get
\ba
\la \Phi^a(x,\th,\bth)\Phi_b(x^\prime,\th^\prime,\bth^\prime) \ra 
&\!\!=\!\!& -\frac{\delta^a_b}{12(2\pi)^2} \,\er^{-\frac{i}{\sqrt{2}}
(\th^m\bth_m-\th^m\bth^\prime_m)\del_-} 
\frac{\delta^{(4)}(\th-\th^\prime)}{(x-x^\prime)^2} \nn \\
&&\times \fr{(4!)^2} \, \er^{+\frac{i}{\sqrt{2}}(\th^m\bth_m-
\th^m\bth^\prime_m)\del_-} \la d^4 \ra \delta^{(4)}(\bth-\bth^\prime) \, ,
\label{super2pt-4}
\ea
where we used the $\delta$-function in (\ref{super2pt-3}) to change
${\th^\prime}^m$ into $\th^m$ in the first exponential. The
exponential factors in (\ref{super2pt-4}) cancel and we finally get
\be
\la \Phi^a(x,\th,\bth)\Phi_b(x^\prime,\th^\prime,\bth^\prime) \ra 
= -\frac{2}{(4!)^3} \,\frac{\delta^a_b}{(2\pi)^2}\, 
\frac{\la d^4 \ra \delta^{(4)}(\th-\th^\prime)
\delta^{(4)}(\bth-\bth^\prime)}{(x-x^\prime)^2} \, .
\ee
in agreement with (\ref{super2pt-fin}).

\section{Useful superspace relations}
\label{app:superspace-identities}

We collect in this appendix various relations used in manipulations of
super Feynman diagrams in light-cone superspace. 

Although $\fr{\del_-}$ is not a differential operator, it can be
``integrated  by parts'' in superspace expressions. For generic
superfields $f(x,\th,\bth)$ and $g(x,\th,\bth)$ we have 
\ba
&&\hsp{-1} \int\dr^{12}z \, f(z)\fr{\del_-}g(z) = 
\int\dr^{12}z\,\frac{\del_-}{\del_-} f(z) \fr{\del_-}g(z) \nn \\
&& \hsp{2.3} = -\int \dr^{12}z\, \fr{\del_-} f(z) 
\frac{\del_-}{\del_-} g(z) = -\int\dr^{12}z\,\fr{\del_-}f(z)g(z) \, .
\label{integrbyparts}
\ea
Using the definition (\ref{chiralder}) of the chiral derivatives,
$d^m$ and $\bar d_m$, and their commutation relation, it is easy to
verify the following identity
\be
\int \dr^{12} z_2 \, \d^{(8)}(\th_1-\th_2) \left[\la d_{(1)}^4\ra
\la\bar d_{(1)}^4\ra \d^{(8)} (\th_1-\th_2) \right] = (4!)^4 \, ,
\label{GRS-1}
\ee
which is used repeatedly to carry out the integrations over the
fermionic coordinates at each interaction vertex in superspace Feynman
diagrams. 

The commutation relation (\ref{dsusyalg}) for the superspace chiral
derivatives implies
\be
\label{eq17}
\overrightarrow{\la \bar d^4\ra \la d^4 \ra \bar d_p \bar d_q} 
= 4!\, \veps_{abpq}\, \overrightarrow{\parm^2 \la \bar d^4\ra d^a d^b}\,,
\ee
\be
\label{eqn17}
\overrightarrow{\la \bar d^4\ra\la d^4\ra\la \bar d^4\ra}
=4(4!)^2\overrightarrow{\parm^4\la\bar d^4\ra}\,.
\ee
The following identity can be verified using the normalisation of
Grassmann integrals in appendix \ref{app:conventions}
\be
\int \dr^4\th\dr^4\bth\,\th^m\th^n\th^p\th^q\bth_m\bth_n\bth_p\bth_q 
= \fr{4!} \, .
\ee

\section{Details of four-point function calculation}
\label{app:calc-details}

\subsection{Diagrams involving cubic vertices}
\label{app:rules-cubic}

As pointed out in section \ref{connect-cubic} contributions to
four-point functions of the $Q^{ij}$ operators cannot be built using
two cubic vertices of the same type (Vertex 3-I in (\ref{3-vertex_1})
or Vertex 3-II in (\ref{3-vertex_2})). This can be seen from a simple
counting of chiral derivatives and fermionic coordinates $\th$ and
$\bth$. 

We start by counting the superficial numbers (or powers) of $d$, $\bar
d$, $\th$ and $\bth$ present in various factors used in constructing a
four point function. 

\begin{figure}[htb]
\begin{center}
\includegraphics[width=0.35\textwidth]{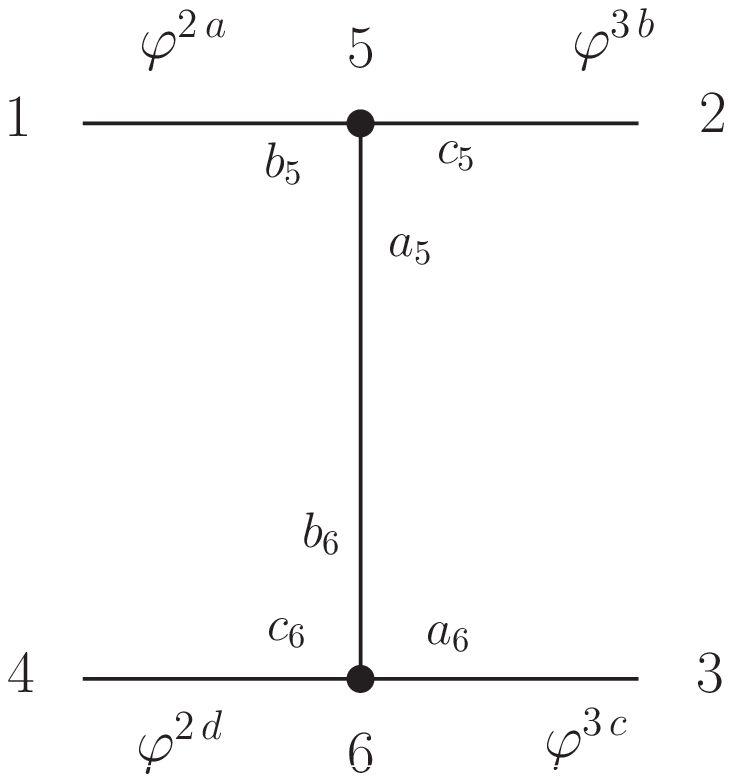}
\end{center}
\caption{}
\label{fig:disc-cubic-2-int}
\end{figure}

\begin{table}[htb]
\begin{center}
\begin{tabular}[b]{|l|c|c|c|c|}
\hline
Structure \rule[-7pt]{0pt}{20pt} & $d$ & $\bar d$ & $\th$ & $\bth$  \\ 
\hline\hline
Propagator & 4 & 0 &  4 & 4  \\ 
\hline
Cubic Vertex 3-I & 0 & 4 & 0 & 0 \\
\hline
Cubic Vertex 3-II & 0 & 8 & 0 & 0 \\
\hline
External $\v$ field in $\calQ$ & 0 & 2 & 0 & 0\\
\hline
\end{tabular}
\caption{Superficial powers of $d$, $\bar d$, $\th$, $\bth$}
\end{center}
\end{table}
\ndt
The superficial numbers (or powers) of various derivatives and
fermionic variables in a four point function as shown in figure
\ref{fig:disc-cubic-2-int}, are presented in table \ref{tab:cubic} for
the three possible cases.

\begin{table}[htb]
\begin{center}
\begin{tabular}[b]{|l|c|c|c|c|c|c|}
\hline
Combination of vertices \rule[-7pt]{0pt}{20pt}
& $d$ & $\bar d$ & $\th$ & $\bth$ & $\dr \th$ & $\dr \bth$ \\ 
\hline\hline
Vertex 3-I and Vertex 3-II & 20 & 20 &  20 & 20 & 8 & 8 \\ 
\hline
Vertex 3-I twice & 20 & 16 & 20 & 20  & 8 & 8\\
\hline
Vertex 3-II twice & 20 & 24 & 20 & 20 & 8 & 8 \\
\hline
\end{tabular}
\caption{Superficial powers of $d$, $\bar d$, $\th$, $\bth$, $\dr
\th$, $\dr \bth$  in a four point function}
\label{tab:cubic}
\end{center}
\end{table}
\ndt
After performing the fermionic integrals in a super Feynman diagram,
we are left with an \textit{equal} number of $\th$'s and
$\bth$'s. Thus when fermionic coordinates are set to zero, a
non-vanishing contribution can only arise if there are \textit{equal}
numbers of $d$'s and $\bar d$'s present to cancel the $\th$'s and
$\bth$'s. Thus, as can be seen from table \ref{tab:cubic}, only the
combination of one vertex of type 3-I and one of type 3-II can produce
a non-zero result, as this is the only way of satisfying the above
criterion. 
We illustrate this argument with the explicit example of a four point
function at one loop constructed using only the first cubic vertex. In
this example, we focus on a  specific contraction, where the legs
labelled by colours $b_5$ and $a_6$ in figure
\ref{fig:disc-cubic-2-int} carry the $\left(\la\bar d^4\ra/\parm^3\right)$
factor from the cubic vertex. We will henceforth suppress numerical
factors, space-time derivatives and all tensor structures in the colour
and flavour indices as they will not be important in the rest of the
argument. The four point function evaluates to
\ba
&& \int \dr^4 \th_5\, \dr^4 \bth_5\, \dr^4 x_5 \int \dr^4 \th_6\, 
\dr^4 \bth_6\, \dr^4 x_6 \nn \\
&& \times \left( \la \bar d^4\ra \la d^4\ra \frac{\d^8_{51}}{x^2_{51}} 
\overleftarrow{\bar d \bar d} \right)\!
\left( \la d^4 \ra \frac{\d^8_{52}}{x^2_{52}} 
\overleftarrow{\bar d \bar d} \right) \!
\left(\la d^4\ra \frac{\d^8_{56}}{x^2_{56}} \right)\! 
\left( \la d^4\ra \frac{\d^8_{64}}{x^2_{64}} 
\overleftarrow{\bar d\bar d}\right)\! 
\left( \la \bar d^4\ra \la d^4\ra \frac{\d^8_{63}}{x^2_{63}}
\overleftarrow{\bar d\bar d}\right)\,.
\ea
Using the relation 
\[ \la \bar d^4 \ra \la d^4\ra \d^8 \overleftarrow{\bar d\bar d} 
\sim \la \bar d^4 \ra \la d^4\ra \bar d\bar d\,\d^8 
\sim \la \bar d^4\ra d d \,\d^8\,, \]
the integrand simplifies to
\[
\left( \la \bar d^4\ra dd \frac{\d^8_{51}}{x^2_{51}} \right)\!
\left( \la d^4 \ra \bar d \bar d\frac{\d^8_{52}}{x^2_{52}} \right)\! 
\left(\la d^4\ra \frac{\d^8_{56}}{x^2_{56}} \right)\! 
\left( \la d^4\ra \bar d \bar d\frac{\d^8_{64}}{x^2_{64}} \right)\! 
\left( \la \bar d^4\ra dd \frac{\d^8_{63}}{x^2_{63}}\right)\,.
\]
We partially integrate $\la d^4\ra$ from the second bracket to the
first bracket, use the simplifying relation $ \la d^4 \ra \la \bar
d^4\ra d d\,\d^8 \sim \la d^4\ra \bar d\bar d\,\d^8$, and move the two
$\bar d$'s in the second bracket to the first bracket as this is the
only term that may survive once we perform all fermionic integrals and
set $\th$'s and $\bth$'s to zero. We first integrate over $\th_5$ and
$\bth_5$ using the free delta function $\d^8_{52}$, and obtain
\ba
&& \int \dr^4 x_5 \left( \bar d \bar d\la d^4\ra \bar d\bar d 
\frac{\d^8_{21}}{x^2_{51}} \right)\!
\left( \frac{1}{x^2_{52}} \right) \nn \\
&& \times \int \dr^4 \th_6\, \dr^4 \bth_6\, \dr^4 x_6 
\left(\la d^4\ra \frac{\d^8_{62}}{x^2_{56}} \right)\! 
\left( \la d^4\ra \bar d \bar d\frac{\d^8_{64}}{x^2_{64}} \right)\! 
\left( \la \bar d^4\ra dd \frac{\d^8_{63}}{x^2_{63}}\right)\,.
\ea
We partially integrate $\la d^4\ra$ from the first bracket to the
third bracket inside the fermionic integral to free up the delta
function $\d^8_{62}$, simplify the combination of the chiral and
anti-chiral derivatives in the third bracket and then perform the
remaining fermionic integrals to obtain,
\be
\int \dr^4 x_5 \int \dr^4 x_6 \left( \bar d \bar d\la d^4\ra 
\bar d\bar d \frac{\d^8_{21}}{x^2_{51}} \right)\!
\left( \frac{1}{x^2_{52}} \right)\! \left(\frac{1}{x^2_{56}} \right)\! 
\left( \la d^4\ra \bar d \bar d\frac{\d^8_{24}}{x^2_{64}} \right)\! 
\left( \la d^4\ra \bar d\bar d \frac{\d^8_{23}}{x^2_{63}}\right)\,.
\ee
We find that the fourth and fifth brackets have an insufficient number
of $\bar d$'s (precisely two each) to cancel the $\th$'s and thus this
expression reduces to zero when we set the external $\th$'s and
$\bth$'s to zero.

If we choose to work with only the second cubic vertex, using similar
manipulations we end up with an insufficient number of $d$'s
(precisely two $d$'s per term in a total of two terms), so that the
expression reduces to zero when we set the fermionic coordinates to
zero.

\vsp{0.3}

\begin{rules} In the topology shown in Figure
\ref{fig:disc-cubic-2-int}, a cubic vertex cannot have component
fields $\v^i$ and $\v^j$ with $i\neq j$, connected to any two of its
legs.
\label{rules-cubic}
\end{rules}

We will show why this is the case through an explicit calculation of a
particular arrangement of the cubic vertices in figure
\ref{fig:disc-cubic-2-int} in which we assume that we have Vertex 3-I
at point $z_5$ and Vertex 3-II at point $z_6$. The calculations for
the other permutations are identical.  There are in total $3!\times
3!$ possible permutations of the cubic vertices with Vertex 3-I
(\ref{3-vertex_1}) at $x_5$ and Vertex 3-II (\ref{3-vertex_2}) at
$x_6$. Thus a total of $3!\times 3!\times 2$ possible Wick
contractions (including the cases when Vertex 3-I is at $x_6$ and
Vertex 3-II at $x_5$). The contractions in figure \ref{fig:disc-cubic-2-int}
give
\ba
I\!&=&\!\! \sigma^{2m_1n_1}\sigma^{3m_2n_2}\sigma^{3m_3n_3}
\sigma^{2m_4n_4}\left(\frac{g^2}{12}\right)
f^{a_5b_5c_5}f^{a_6b_6c_6}k^5 \nn \\
&& \times \!\!\int_{5,6}\!\!  \left(\!\la d^4\ra 
\frac{\d^8_{51}}{x^2_{51}}\overleftarrow{\bar d_{n_1}\bar d_{m_1}}
\d^{ab_5}\!\right)\!\!
\left(\!\bar \del \la d^4\ra \frac{\d^8_{52}}{x^2_{52}}
\overleftarrow{\bar d_{n_2}\bar d_{m_2}}\d^{bc_5}\!\right)\!\!
\left(\!\frac{\la \bar d^4\ra}{\parm^3}\la d^4\ra 
\frac{\d^8_{56}}{x^2_{56}}\overleftarrow{
\frac{\la \bar d^4\ra}{\parm^2}}\d^{a_5b_6}\!\right)\!\!\! \nn \\
&& \times \left(\!\fr{\parm}\la d^4\ra \frac{\d^8_{63}}{x^2_{63}}
\overleftarrow{\bar d_{n_3}\bar d_{m_3}}\d^{ca_6}\!\right)\!\!
\left(\! \frac{\del \la \bar d^4\ra}{\parm^2}\la d^4\ra 
\frac{\d^8_{64}}{x^2_{64}}\overleftarrow{
\bar d_{n_4}\bar d_{m_4}}\d^{dc_6}\!\right) \\
&=& \!\! \sigma^{2m_1n_1}\sigma^{3m_2n_2}
\sigma^{3m_3n_3}\sigma^{2m_4n_4}\left(
\frac{g^2}{12}\right)f^{a_5b_5c_5}f^{a_6b_6c_6}k^5\nn \\
&& \times \!\!\int_{5,6}\!\! \left(\!\la d^4\ra 
\frac{\d^8_{51}}{x^2_{51}}\overleftarrow{\bar d_{n_1}\bar d_{m_1}}
\d^{ab_5}\!\right)\!\!
\left(\!\bar \del \la d^4\ra \frac{\d^8_{52}}{x^2_{52}}
\overleftarrow{\bar d_{n_2}\bar d_{m_2}}\d^{bc_5}\!\right)\!\!
\left(4(4!)^2\parm \la\bar d^4\ra \frac{\d^8_{56}}{x^2_{56}}
\overleftarrow{\fr{\parm^2}}\d^{a_5b_6}\!\right)\nn \\
&& \times \left(\!\fr{\parm}\la d^4\ra \frac{\d^8_{63}}{x^2_{63}}
\overleftarrow{\bar d_{n_3}\bar d_{m_3}}\d^{ca_6}\!\right)\!\!
\left(\! 4!\veps_{abn_4m_4}\del \la \bar d^4\ra d^ad^b
\frac{\d^8_{64}}{x^2_{64}}\d^{dc_6}\!\right)\,,
\label{cubic-rule-int}
\ea
where we used (\ref{eqn17}). 

The second term inside the integral in (\ref{cubic-rule-int}) can be
rewritten as $\left(\!\bar \del \la d^4\ra\bar d_{n_2}\bar d_{m_2}
\frac{\d^8_{52}}{x^2_{52}}\d^{bc_5}\!\right)$, and now we can
partially integrate $\la d^4\ra$ entirely to the term containing
$\d^8_{56}$. Now, the derivatives $\bar d_{n_2}$ and $\bar d_{m_2}$
have two possible destinations on partial integration, but only when
both move to the term containing $\d^8_{51}$ can we hope to get a
non-zero contribution in the limit of fermionic coordinates going to
zero. Thus $I$ simplifies to
\ba
&&\hsp{-0.6}4(4!)^3\sigma^{2m_1n_1}\sigma^{3m_2n_2}
\sigma^{3m_3n_3}\sigma^{2m_4n_4}\left(\frac{g^2}{12}\right)
f^{a_5b_5c_5}f^{a_6b_6c_6}k^5 
\d^{ab_5}\d^{bc_5}\d^{a_5b_6}\d^{ca_6}\d^{dc_6}
\veps_{rsn_4m_4} \nn \\
&& \hsp{-0.6}\times \!\!\!\int_{5,6}\!\!\! \left(\!\bar d_{n_2}
\bar d_{m_2} \la d^4\ra \frac{\d^8_{51}}{x^2_{51}}
\overleftarrow{\bar d_{n_1}\bar d_{m_1}}\!\right)\!\!\!
\left(\!\bar \del \frac{\d^8_{52}}{x^2_{52}}\!\right)\!\!\!
\left(\!\parm \la d^4\ra\la\bar d^4\ra \frac{\d^8_{56}}{x^2_{56}}
\overleftarrow{\fr{\parm^2}}\!\right)\!\!\!
\left(\!\frac{\la d^4\ra}{\parm} \frac{\d^8_{63}}{x^2_{63}}
\overleftarrow{\bar d_{n_3}\bar d_{m_3}}\!\right)\!\!\!
\left(\! \del \la \bar d^4\ra d^rd^s\frac{\d^8_{64}}{x^2_{64}}\!
\right)\nn\\
&&\hsp{-0.6}=4(4!)^3\sigma^{2m_1n_1}\sigma^{3m_2n_2}
\sigma^{3m_3n_3}\sigma^{2m_4n_4}\left(\frac{g^2}{12}\right)
f^{a_5b_5c_5}f^{a_6b_6c_6}k^5 
\d^{ab_5}\d^{bc_5}\d^{a_5b_6}\d^{ca_6}\d^{dc_6}
\veps_{rsn_4m_4} \nn \\
&&\hsp{-0.6}\times \!\!\!\int\!\! \dr^4 x_5\!\!
\left(\!\bar \del \frac{1}{x^2_{52}}\!\right)\!\!\!
\left(\!\bar d_{n_2}\bar d_{m_2} \la d^4\ra 
\frac{\d^8_{21}}{x^2_{51}}\overleftarrow{\bar d_{n_1}
\bar d_{m_1}}\!\right)\!\!\!\int_{6}\!\!\! \left(\!\parm \la d^4\ra\la
\bar d^4\ra \frac{\d^8_{26}}{x^2_{56}}
\overleftarrow{\fr{\parm^2}}\!\right)\!\!\!\left(\!
\frac{\la d^4\ra}{\parm} \frac{\d^8_{63}}{x^2_{63}}
\overleftarrow{\bar d_{n_3}\bar d_{m_3}}\!\right)\!\!\!
\left(\! \del \la \bar d^4\ra d^rd^s\frac{\d^8_{64}}{x^2_{64}}\!
\right)\nn\\
&&\hsp{-0.6}=4(4!)^6\sigma^{2m_1n_1}\sigma^{3m_2n_2}
\sigma^{3m_3n_3}\sigma^{2m_4n_4}\left(\frac{g^2}{12}\right)
f^{a_5b_5c_5}f^{a_6b_6c_6}k^5 
\d^{ab_5}\d^{bc_5}\d^{a_5b_6}\d^{ca_6}\d^{dc_6}
\veps_{rsn_4m_4} \nn \\
&&\hsp{-0.6}\times \veps_{n_2m_2n_1m_1}
\!\!\!\int\!\! \dr^4 x_5\!\!\left(\!\bar \del \frac{1}{x^2_{52}}\!\right)
\!\!\!\left(\frac{1}{x^2_{51}}\!\right)\times J\, , \nn
\ea
where 
\[
\displaystyle{J}
\equiv \lim_{\th,\bth\rightarrow 0}\int_{6}\!\left(\!\parm \la
d^4\ra\la\bar d^4\ra
\frac{\d^8_{26}}{x^2_{56}}\overleftarrow{\fr{\parm^2}}\!\right)\!\!
\left(\!\frac{\la
d^4\ra}{\parm} \frac{\d^8_{63}}{x^2_{63}}\overleftarrow{\bar
d_{n_3}\bar d_{m_3}}\!\right)\!\!\left(\! \del \la \bar d^4\ra
d^rd^s\frac{\d^8_{64}}{x^2_{64}}\!\right) \, .
\] 
In the limit $\th,\bth\! \rightarrow\! 0$ the above expression
vanishes since
$\sigma^{2m_1n_1}\sigma^{3m_2n_2}\veps_{n_2m_2n_1m_1}=0$.

In general, attaching $\v^i$ and $\v^j$ to two legs of a cubic vertex
will result in a factor of
$\sigma^{i\,mn}\sigma^{j\,pq}\veps_{mnpq}=8\d^{ij}$. Thus such an
arrangement with $i\neq j$ does not contribute.

This result can be understood in terms of component fields. The only
cubic vertices involving two scalar fields in the $\calN=4$ action --
in any gauge, including the light-cone gauge -- are the minimal
coupling to the gauge field. Since the latter is a flavour singlet,
the interaction cannot change the flavour index carried by the scalar
field.

\subsection{Diagrams involving quartic vertices}
\label{app:rules-quartic}

\begin{figure}[htb]
\begin{center}
\includegraphics[width=0.35\textwidth]{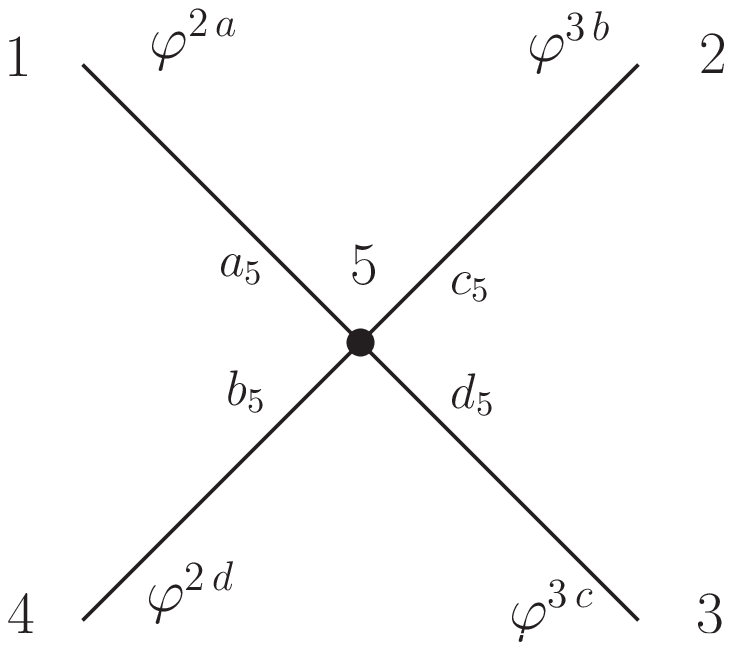}
\end{center}
\caption{}
\label{fig:disc-quartic-int}
\end{figure}

\begin{rules} 
In the topology shown in Figure \ref{fig:disc-quartic-int}, component
fields $\v^i$ and $\v^j$ with flavour $i\neq j$, cannot simultaneously
attach to those legs of the quartic vertex which are both chiral
fields, or both anti-chiral fields~\footnote{Here we use the term
``anti-chiral'' field to refer to superfields associated with legs in
a diagram carrying a $\la\bar d^4\ra/\del_-^2$ factor. These were
originally $\bar\Phi$'s before use of (\ref{Phibar}).}.
\label{rules-quartic}
\end{rules}

For a four point function constructed using Vertex 4-I
(\ref{4-vertex_1}), if the leg with colour index $a_5$ (chiral field)
is connected with the external field $\v^i$, and the leg with colour
index $b_5$ (chiral field) with the field $\v^j$, we get a factor of
$\sigma^{i\,mn}\sigma^{j\,pq}\veps_{mnpq} = 8\d^{ij}$ when evaluating
the correlation function. The same thing happens with legs carrying
colour indices $c_5$ and $d_5$ (anti-chiral fields) connected with
external fields $\v^i$ and $\v^j$. For Vertex 4-II (\ref{4-vertex_2}),
if the leg with colour index $a_5$ (chiral field) is connected with
$\v^i$ and the leg with index $c_5$ (chiral field) with $\v^j$, we get
a factor of $\sigma^{i\,mn}\sigma^{j\,pq}\veps_{mnpq} = 8\d^{ij}$. The
same happens with legs carrying colour indices $d_5$ and $b_5$
(anti-chiral fields). Thus for such arrangements with $i \neq j$, the
contraction vanishes.

This rule is verified by evaluating each permutation of the
interaction vertex in Figure \ref{fig:disc-quartic-int} and performing
manipulations similar to those done in section \ref{app:rules-cubic},
for both Vertex 3-I and Vertex 3-II.

\vsp{0.3}

The only non-zero contributions to $G_4^{(H)}(x_1,\ldots,x_4)$ at one
loop come from diagrams involving a quartic vertex of type 4-II. As
explained in section \ref{quart-conn} there are various inequivalent
Wick contractions to consider and we analyse them in detail below. We
begin with \\
\hsp{0.5}\raisebox{-20pt}{
\includegraphics[width=0.15\textwidth]{V4-abcd.eps}} 
$\equiv$ $V_4[a_5,b_5,c_5,d_5]$
\ba
&& \hsp{-0.1} = \int_5 \delta^{aa_5}\delta^{bb_5}\delta^{cc_5}
\delta^{dd_5} \left( \bar d_{p_1} \bar d_{q_1} \la d^4\ra 
\frac{\delta^8_{15}}{x^2_{15}} \right) \!\!
\left( \frac{\la \bar d^4 \ra}{\parm^2} \la d^4 \ra 
\frac{\delta^8_{52}}{x^2_{52}} \overleftarrow{\bar d_{q_2} 
\bar d_{p_2}} \right) \nn \\ 
&& \hsp{3} \times \left( \bar d_{p_3} \bar d_{q_3} 
\la d^4\ra \frac{\delta^8_{35}}{x^2_{35}} \right)\!\!
\left( \frac{\la \bar d^4 \ra}{\parm^2} \la d^4 \ra 
\frac{\d^8_{54}}{x^2_{54}} \overleftarrow{\bar d_{q_4} \bar d_{p_4}} 
\right) \nn \\
&& \hsp{-0.1}\propto \veps_{p_1q_1p_3q_3}\,. \nn
\ea
Product with the common part $E_4[a_5,b_5,c_5,d_5]$ in
(\ref{common-1}) results in the contraction
$\sigma^{2p_1q_1}\sigma^{3p_3q_3} \veps_{p_1q_1p_3q_3}\!~=~\!0$. The
reason why $V_4[a_5,b_5,c_5,d_5]$ leads to this contraction is
explained under Rule \ref{rules-quartic} above. \\
\hsp{0.5}\raisebox{-20pt}{
\includegraphics[width=0.15\textwidth]{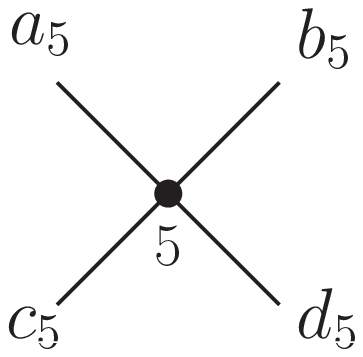}} 
$\equiv$ $V_4[a_5,b_5,d_5,c_5]$
\ba
&&=\int_5 \d^{a_5a}\d^{b_5b}\d^{c_5d}\d^{d_5c} \left( 
\bar d_{p_1} \bar d_{q_1} \la d^4\ra 
\frac{\delta^8_{15}}{x^2_{15}} \right)\!\! \left( 
\frac{\la \bar d^4 \ra}{\parm^2} \la d^4 \ra 
\frac{\d^8_{52}}{x^2_{52}} \overleftarrow{\bar d_{q_2} \bar d_{p_2}} 
\right) \!\!\left( \bar d_{p_3} \bar d_{q_3} \la d^4\ra 
\frac{\d^8_{35}}{x^2_{35}} 
\overleftarrow{\frac{\la \bar d^4 \ra}{\parm^2}} \right)\nn \\
&& \qquad \qquad  \qquad \qquad \times \left(\la d^4 \ra 
\frac{\d^8_{54}}{x^2_{54}} \overleftarrow{\bar d_{q_4} \bar d_{p_4}} 
\right) \nn \\
&& = \int_5 \d^{a_5a}\d^{b_5b}\d^{c_5d}\d^{d_5c} \left(
\la d^4\ra \bar d_{q_1} \bar d_{p_1}  \frac{\d^8_{51}}{x^2_{51}} 
\right) \!\!\left( 4! \veps_{rsq_2p_2}\la \bar d^4 \ra d^r d^s 
\frac{\d^8_{52}}{x^2_{52}} \right) \!\! \left( 4!\veps_{uvq_3p_3} 
\la \bar d^4\ra d^u d^v \frac{\d^8_{53}}{x^2_{53}} \right)\nn \\
&& \qquad \qquad  \qquad \qquad \times \left(\la d^4 \ra 
\bar d_{q_4} \bar d_{p_4} \frac{\d^8_{54}}{x^2_{54}} \right)\,,
\ea
where we used (\ref{eq17}).

We now use the following rule for partially integrating $\la d^4 \ra$
to a product of two terms (disregarding the cases where both the terms
are not acted upon by two $d$'s each), 
\be
\int \dr^4 \theta (\la d^4 \ra F)(GH) = 6\, \veps_{m_1n_1m_2n_2} 
\int \dr^4 \theta\, F(d^{m_1} d^{n_1} G) (d^{m_2} d^{n_2} H)\,,
\ee
and simplify $V_4[a_5,b_5,d_5,c_5]$ to
\ba
&& \hsp{-1}6 \!\!\int_5 d^{a_5a}\d^{b_5b}\d^{c_5d}\d^{d_5c} 
\left( \frac{\d^8_{51}}{x^2_{51}}\right)\!\!\left( 4! 
\veps_{rsq_2p_2}d^{m_1} d^{n_1}\la \bar d^4 \ra d^r d^s 
\frac{\d^8_{52}}{x^2_{52}} \right)\!\!\!  \nn \\
&& \times \left( 4!\veps_{uvq_3p_3} 
d^{m_2} d^{n_2}\la \bar d^4\ra d^u d^v \frac{\d^8_{53}}{x^2_{53}} 
\right)\!\!\left(\bar d_{q_1} \bar d_{p_1}\la d^4 \ra \bar d_{q_4} 
\bar d_{p_4} \frac{\d^8_{54}}{x^2_{54}} \right)  
\veps_{m_1n_1m_2n_2}\,.
\label{ABDC-1}
\ea
$V_4[a_5,b_5,d_5,c_5]$ as written in (\ref{ABDC-1}) simplifies to
\ba
&& \hsp{-1}6(4!)^2 \!\!\!\int\!\! \dr^4 x_5\, 
d^{a_5a}\d^{b_5b}\d^{c_5d}\d^{d_5c} 
\!\!\left( \frac{\veps_{m_1n_1m_2n_2}}{x^2_{51}} \right)\!\! \left( 
\frac{(4!)^3 \veps_{rsq_2p_2} \veps^{m_1n_1rs}}{x^2_{52}}\right) \nn \\
&&\hsp{1}\times\left(\frac{(4!)^3\veps_{uvq_3p_3}
\veps^{m_2n_2uv}}{x^2_{53}}\right)\!\!
\left(\frac{(4!)^3\veps_{q_1p_1q_4p_4}}{x^2_{54}}\right),
\label{ABDC-2}
\ea
in the limit $\th,\bth \rightarrow 0$. Using the following property of
the Levi-Civita symbol
\be
\veps_{m_1n_1m_2n_2}\:\veps^{m_1n_1pq} = 
2\left(\d^p_{m_2}\d^q_{n_2} - \d^p_{n_2}\d^q_{m_2}\right)\,,
\ee
we simplify
\be
\left(\veps_{m_1n_1m_2n_2}\:\veps^{m_1n_1rs}\right)
\veps_{rsq_2p_2} \left(\veps_{uvq_3p_3}\veps^{m_2n_2uv}\right)
= 4\:\veps_{m_2n_2q_2p_2} \left(\veps_{uvq_3p_3}
\veps^{m_2n_2uv}\right)=16\:\veps_{p_3q_3p_2q_2}\,.
\ee
Thus $V_4[a_5,b_5,d_5,c_5]$ (\ref{ABDC-2}) simplifies to
\be
16 \times 6 \times (4!)^{11} \times 
d^{a_5a}\d^{b_5b}\d^{c_5d}\d^{d_5c} \times 
\veps_{p_3q_3p_2q_2}  \veps_{p_1q_1p_4q_4} 
\int \dr^4 x_5\, \fr{x^2_{51}x^2_{52}x^2_{53}x^2_{54}}\,.
\ee
Substituting
\be
k=(-1).2.\fr{(4!)^3}\fr{(2\pi)^2}\,, \qquad 
T(\sigma)\veps\veps\veps\veps=2^{12} \,,
\ee
we obtain the final expression for $V_4[a_5,b_5,d_5,c_5]$ times the
common part (\ref{common-2}) as
\be
-g^2f^{eab}f^{eab} \frac{1}{(2\pi)^{12}} \fr{x^2_{14}x^2_{23}}
\int \dr^4 x_5 \fr{x^2_{51}x^2_{52}x^2_{53}x^2_{54}}\,.
\ee
All permutations of the arguments in $V_4[a_5,b_5,d_5,c_5]$ of the
form $[e_1,g_1,e_2,g_2]$ where $e_i\in\{a_5,c_5\}, g_i\in \{b_5,d_5\}$ or
$e_i\in\{b_5,d_5\}, g_i\in\{a_5,c_5\}$, $i=1,2$, will have a non-zero
contribution. The reason is explained under Rule \ref{rules-quartic}
above.

From the structure of Vertex 4-II (\ref{4-vertex_2}), it is easy to
see that
\ba
V_4[a_5,b_5,d_5,c_5] = V_4[a_5,d_5,b_5,c_5] = 
V_4[c_5,b_5,d_5,a_5] = V_4[c_5,d_5,b_5,a_5] \,\,\, && \nn \\
= V_4[b_5,a_5,c_5,d_5] = V_4[b_5,c_5,a_5,d_5] = 
V_4[d_5,a_5,c_5,d_5] = V_4[d_5,c_5,a_5,b_5] \, . &&  \nn
\ea

\vsp{1}

\end{document}